\documentclass[lettersize,journal]{IEEEtran}
\usepackage{amsmath,amsfonts}
\usepackage{algorithmic}
\usepackage{algorithm}
\usepackage{array}
\usepackage[caption=false,font=normalsize,labelformat=simple]{subfig}

\usepackage[skins, breakable, listings, theorems]{tcolorbox}

\usepackage{booktabs}
\usepackage{textcomp}
\usepackage{stfloats}
\usepackage{url}
\usepackage{hyperref}
\usepackage{verbatim}
\usepackage{graphicx}
\usepackage{cite}
\usepackage{amsmath,amsfonts}
\usepackage{textcomp}
\usepackage{verbatim}
\usepackage{xcolor}
\usepackage{soul}
\usepackage{comment}

\begin{document}

\title{This work has been submitted to the IEEE for possible publication. Copyright may be transferred without notice, after which this version may no longer be accessible.
\vspace{2 cm}

Teaching LLMs to See and Guide: Context-Aware Real-Time Assistance in Augmented Reality}

\author{Mahya Qorbani\IEEEauthorrefmark{1},
        Kamran Paynabar\IEEEauthorrefmark{1},
        and Mohsen Moghaddam\IEEEauthorrefmark{1}\IEEEauthorrefmark{2}%

\IEEEauthorblockA{\IEEEauthorrefmark{1} H. Milton Stewart School of Industrial and Systems Engineering, Georgia Institute of Technology, Atlanta, GA, USA}\\
\IEEEauthorblockA{\IEEEauthorrefmark{2} George W. Woodruff School of Mechanical Engineering, Georgia Institute of Technology, Atlanta, GA, USA}

    \thanks{This paper was produced by the IEEE Publication Technology Group. They are in Piscataway, NJ.}

\thanks{Mahya Qorbani and Kamran Paynabar are with the H. Milton Stewart School of Industrial and Systems Engineering, Georgia Institute of Technology, Atlanta, GA, USA. (e-mails: mqorbani3@gatech.edu, kpaynabar3@gatech.edu)}%
\thanks{Mohsen Moghaddam is with the H. Milton Stewart School of Industrial and Systems Engineering and the George W. Woodruff School of Mechanical Engineering, Georgia Institute of Technology, Atlanta, GA, USA (e-mail: mohsen.moghaddam@gatech.edu).}%
\thanks{This work was supported by the National Science Foundation under Grant No. 2128743. Any opinions, findings, or conclusions expressed in this material are those of the authors and do not reflect the views of the NSF.}%
\thanks{*Corresponding author: Mohsen Moghaddam (e-mail address: mohsen.moghaddam@gatech.edu).}}

\markboth{IEEE TRANSACTIONS ON SYSTEMS, MAN, AND CYBERNETICS: SYSTEMS,~Vol.~X, No.~X, X~20XX}%
{Qorbani \MakeLowercase{\textit{et al.}}: Context-Aware LLM Assistant for AR Real-Time Task Guidance}

\maketitle

\begin{abstract}
The growing adoption of augmented and virtual reality (AR/VR) technologies in industrial training and on-the-job assistance has created new opportunities for intelligent, context-aware support systems. As workers perform complex tasks guided by AR/VR, these devices capture rich streams of multimodal data—including gaze, hand actions, and task progression—that can reveal user intent and task state in real time.  Leveraging this information effectively remains a major challenge. In this work, we present a context-aware LLM assistant that integrates diverse data modalities, such as task description, hand actions, and dialogue history, into a unified framework for real-time question answering. To systematically study how context influences performance, we introduce an incremental prompting framework, where each model version receives progressively richer contextual inputs. Using the HoloAssist dataset, which records AR-guided task executions, we evaluate how each modality contributes to the assistant’s effectiveness. Our experiments show that incorporating multimodal context significantly improves the accuracy and relevance of responses. These findings highlight the potential of LLM-driven multimodal integration to enable adaptive, real-time, and intuitive assistance for AR/VR-based industrial training and assistance.
\end{abstract}


\section{Introduction}
\IEEEPARstart{A}{ugmented Reality} (AR) is rapidly gaining traction in industrial training and on-the-job assistance, where workers perform complex, safety-critical tasks that demand precision and adaptability \cite{Butaslac2023, Erkoyuncu2017, Moghaddam2021, Egger2020}. By overlaying digital information directly within the user’s field of view, AR enhances productivity, reduces cognitive load, accelerates skill acquisition, and minimizes operational errors across domains ranging from industrial maintenance to surgical procedures. As AR and virtual reality (VR) technologies become increasingly embedded in workplaces, they generate rich streams of multimodal data, including gaze, hand motions, dialogue, and task progression, that can reveal valuable insights into user intent and task context in real time. Leveraging this information effectively can transform AR from a passive display tool into an intelligent, adaptive assistant capable of understanding what the user is doing and anticipating what they need next \cite{Yoo2024, Yang2025, Arakawa2024, Mucha2024}. At the same time, recent advances in large language models (LLMs) such as GPT have revolutionized machine understanding and interaction. These models excel at reasoning across modalities, adapting to context, and engaging in natural, task-specific dialogue \cite{Shim2025, Cheng2025, Urooj2020, Wang2024, Althaf2025}. Their success in question answering, content summarization, and instructional guidance highlights their potential to support real-time decision-making and knowledge transfer in AR environments.

For AR task guidance to achieve its full potential, it must be deeply intertwined with the surrounding context. Current AR guidance systems often rely on pre-programmed, rigid instructions that struggle to adapt to the inherent variability of real-world scenarios, the user's unique perspective, or unexpected deviations \cite{Wang2025}. This limitation can lead to user frustration, errors, and reduced efficiency, hindering the full potential of AR-assisted workflows. Integrating LLMs into AR systems offers a promising path toward intelligent, context-aware assistance \cite{Asadi2024, Li2025, Javaheri2024}. These powerful models offer a compelling opportunity to overcome the limitations of traditional AR guidance by enabling more intuitive, adaptive, and human-like assistance. By processing and understanding user queries and environmental cues in natural language, LLMs can potentially provide dynamic and contextually relevant guidance, much like a human expert would offer. Integrating LLMs within AR systems offers the compelling prospect of creating context-aware, real-time assistants that not only respond accurately to user queries but proactively deliver guidance tailored precisely to ongoing tasks and environmental conditions. Understanding the user's current situation, the objects they are interacting with, and their progress in the task is crucial for providing timely and relevant instructions.

To the best of the authors' knowledge, no existing research leverages multiple data modalities to comprehensively capture context and deliver real-time guidance or respond dynamically to user queries in AR environments. Prior studies have largely relied on predefined instructions activated at fixed times or employed limited functions where LLMs merely select which function to trigger, without deeply integrating diverse AR-specific data sources such as egocentric video, dialogues, or hand actions. This work addresses this gap by developing a context-aware LLM assistant that seamlessly integrates various modalities to offer dynamic, real-time AR task guidance tailored precisely to users' ongoing actions and environmental conditions. The contributions of this work are as follows:
\begin{itemize}

\item We propose a context-aware framework for AR-based task guidance that integrates LLMs with multimodal AR sensor inputs, such as hand actions, egocentric video, and speech, to deliver adaptive, human-like, and real-time assistance. This integration enables contextually-aware guidance that dynamically adapts to the user’s ongoing actions and environment.

\item We introduce an incremental prompting framework that systematically evaluates how progressively richer contextual inputs influence model performance. This setup enables a systematic evaluation of how varying levels of contextual information affect the model’s ability to provide accurate guidance and respond to user queries. Additionally, we conduct two ablation studies to assess the individual and combined contributions of each modality, offering practical insights for modality selection under resource constraints.

\item We validate the proposed system using the HoloAssist dataset \cite{Wang2023}, assessing performance through lexical metrics, LLM-as-a-judge scoring, and human evaluations. The results demonstrate the model’s strong reasoning and instructional capabilities, confirming its potential as an intelligent AR task instructor. Moreover, we show that the LLM-as-a-judge framework closely aligns with human assessments, highlighting its value as a scalable and cost-effective alternative to manual evaluation.

\end{itemize}
Section~\ref{related work} reviews related literature on AR-based task guidance, multimodal learning, and LLM-assisted interaction systems in AR/VR environments. Section~\ref{methodology} presents the proposed context-aware framework, including its multimodal data integration and incremental prompting design. Section~\ref{sec: experiment} describes the experimental setup and outlines the evaluation procedures. Section~\ref{results} reports the experimental findings, while Section~\ref{discussion} discusses their broader implications and limitations. Section~\ref{conclusions} concludes the paper and outlines directions for future research.

\section{RELATED WORK}
\label{related work}

This section surveys related work on multimodal context modeling, LLMs integration for adaptive, real-time assistance, and AR/VR task guidance.

\subsection{Context-Aware and Memory-Augmented LLMs}
LLMs have shown strong performance in tasks such as conversation, summarization, and question answering. Their success relies heavily on access to relevant context, with both the amount of context available and the effectiveness of context retrieval playing critical roles in enhancing their performance. To address these challenges, we discuss advances in context retrieval and memory structuring techniques, and subsequently examine how prompt engineering approaches enhance LLM performance using retrieved context.

Several systems store past interactions as vector embeddings and retrieve relevant information using semantic similarity search.
For example, \cite{Wu2025} retrieves memory using a combination of semantic search and structured indexing, depending on whether session-based or fact-based memory is needed.
Other approaches leverage tree structures to organize memory hierarchically, improving retrieval efficiency. \cite{Rezazadeh2025b} represents memory as a dynamic tree, where each node contains aggregated textual content, semantic embeddings, and parent-child relationships. In this structure, root nodes store abstract knowledge, while leaf nodes capture more specific details, and memory is updated dynamically as new information arrives.
Similarly,\cite{Aadhithya2024} introduces a memory agent that employs query-conditioned tree traversal to dynamically fetch the most relevant historical data, formulating the retrieval process as an optimal traversal problem instead of relying on flat memory storage. 

Some studies explore graph-based memory representations. \cite{Wilcock2024} discusses how retrieval-augmented generation (RAG) combined with graph databases enhances the accuracy, naturalness, and trustworthiness of dialogue systems. In \cite{Rasal2024}, a temporal graph database tracks conversational history and evolving user preferences, while multiple LLMs collaborate through an orchestration engine that iteratively refines responses based on additional context retrieval. \cite{Alonso2024} proposes a hybrid retrieval approach that combines semantic vector search, structured tabular search (chain-of-tables), and query disambiguation to further enhance long-term memory retrieval in conversational agents. In task-specific domains, \cite{Shim2025} constructs task-centric knowledge graphs from manufacturing documents using LLMs, improving both information extraction and knowledge graph-based question answering for domain experts.

\subsection{Prompt Engineering and Domain Adaptation}
Beyond retrieval mechanisms, prompt engineering strategies are essential for helping LLMs utilize retrieved context effectively during reasoning and response generation. \cite{Pan2025} demonstrates that prompt compression techniques can serve as an effective denoising method to enhance memory retrieval performance. Similarly, \cite{Aadhithya2024} leverages GPT-based models for knowledge aggregation and retrieval, relying solely on prompt engineering to optimize memory selection without requiring additional training. \cite{Lee2023} proposes a modular prompted chatbot, which enables open-domain conversations by combining modular prompting with specialized components such as a clarifier, memory processor (using Dense Passage Retriever and Chain-of-Thought reasoning), utterance generator, and dialogue summarizer. \cite{Wang20241} introduces a recursive summarization strategy using prompt engineering to integrate new information, maintaining long-term context without expanding input length, thus enhancing coherence, fluency, and persona consistency. \cite{Chen2024} utilizes prompt engineering extensively for dataset creation, training, memory compression, and generating memory-grounded responses during conversation. \cite{Maharana2024} combines RAG and prompt engineering, designing prompts that help the model recall user preferences, summarize past interactions, and withstand adversarial memory tests. This enables long-term conversational memory without significantly increasing context length. Similarly, \cite{Hu2024} presents HIAGENT, which structures memory by prompting the LLM to generate subgoals, summarize key actions, and retrieve past subgoals as needed, all without fine-tuning.

Prompt engineering has also been adapted to domain-specific applications. \cite{Zhong2025} introduces TextileBot, a domain-specific voice agent built entirely through prompt engineering strategies without any model fine-tuning. Using taxonomy-based prompting and refined templates (covering identity, behavior, and step-by-step reasoning), the system enables accurate, coherent, multi-turn conversations within a specialized domain. Moreover, \cite{Ravichander2024} shows that answering privacy questions about policies requires both taxonomic reasoning (linking related terms) and regulatory reasoning (handling missing disclosures). They demonstrate that by incorporating structured taxonomies and carefully designed prompts, AI models such as GPT-4 can significantly improve their accuracy on complex privacy question-answering tasks. Additional domain-specific applications include recent works in medical imaging, accessibility, robotics, academic support, technology readiness evaluation, and AR application creation \cite{Makram2025, Salous2025, Chen2025b, Chen2025c, Pabon2025, Betancourt2025, Zhu2025}.

\subsection{Multimodal AI Assistants in AR/VR Environments}
Recently, the input to context-aware LLMs has expanded beyond plain text to include a wide range of data modalities such as tables, figures, speech, images, and video streams \cite{Shim2025, Cheng2025, Urooj2020, Wang2024, Althaf2025}. To support adaptive real-time interaction in immersive environments like AR and VR, systems must leverage these multimodal inputs to deliver context-aware responses. Several recent works have begun exploring this space. For example, \cite{Mucha2024} presents a system where egocentric video from smart AR glasses is analyzed using object detection and optical character recognition to extract text, which is then processed by an LLM to answer user queries, transforming ordinary glasses into intelligent reading assistants. Similarly, \cite{Yang2025} introduces SocialMind, an LLM-based assistive system for AR glasses that uses multimodal sensor data to infer social and contextual cues and assist users in live social interactions without disrupting the natural flow of conversations. \cite{Arakawa2024} proposes PrISM-Q\&A, a smartwatch-based voice assistant for procedural tasks. It uses audio and motion sensor data to estimate the user’s current step through human activity recognition, combining this with the user’s query to enable step-aware LLM-driven assistance. \cite{Lee2024} introduces GazePointAR, a context-aware multimodal voice assistant for wearable AR that leverages eye gaze, pointing gestures, and conversation history to resolve ambiguous pronouns in user queries. 

In the VR domain, \cite{Shao2025} presents CONDA, which reconstructs a user’s home in VR and integrates spatial layout, style, and personal preferences as input to an LLM for furniture recommendation and explanation generation. In mixed reality (MR), \cite{Asadi2024} introduces the Action Sandbox Workspace, where LLMs receive structured prompts representing user goals, spatial context, and available actions. The LLM outputs recommended actions and justifications tailored to the user’s current context and intention. Likewise, \cite{Li2025} presents a framework where GPT-4 Turbo generates real-time, environment-aware responses in VR role-play scenarios using the most recent structured, text-based representations of spatial and interaction context—without using vision. Finally, \cite{Javaheri2024} describes ARAS, an AR-based surgical assistance system that enables intuitive, hands-free interaction through voice commands processed by GPT-3.5 Turbo. The LLM interprets the surgeon’s intent and maps commands to system functions, supporting both preoperative planning and intraoperative guidance in sterile, high-stress surgical environments. 
\\

\subsection{Knowledge Gaps}

While these systems represent exciting progress, they still come with notable limitations. Most depend on brief, recent context and lack the ability to maintain a deeper or longer-term understanding across a full interaction. In many cases, the prompts they use are manually designed or tailored to specific tasks, which makes them fragile and hard to generalize. Another issue is that most systems only work with a narrow range of input types, like speech or video, while overlooking other important signals such as hand movements, eye gaze, or body posture, all of which are providing rich context in immersive AR/VR settings. Lastly, these systems are often built for a single, well-defined purpose, like surgery or social assistance, making it difficult to adapt them to broader, more flexible use cases in real-time environments. These limitations highlight the need for systems that can reason over richer multimodal context, maintain long-term understanding, and generalize across diverse tasks. In this work, we address these challenges by developing a context-aware LLM framework that integrates multiple modalities and employs structured prompting to enable adaptive, real-time guidance across a wide range of task types.

\section{Methodology}
\label{methodology}
The primary objective of this study is to develop and evaluate a framework for real-time user assistance within AR/VR environments. Specifically, we focus on enabling a context-aware LLM assistant that can provide adaptive guidance as users perform complex, multimodal tasks. When users encounter uncertainty or require clarification, the assistant leverages contextual information from the AR/VR environment to deliver timely and relevant support. This section presents the methodology for integrating and configuring the LLM to operate as an interactive, context-aware assistant. The proposed framework is device-agnostic and can be implemented on any AR/VR platform capable of capturing multimodal inputs—including visual (RGB video), speech (dialogue), and sensor-based data such as hand actions, gaze tracking, and inertial measurements. Section~\ref{sec: experiment} details the preprocessing and synchronization of these data modalities to ensure consistency and suitability for input to the LLM assistant.

We adopt an incremental prompting strategy, where the LLM progressively receives relevant contextual information derived from real-time sensors feedback, dialogue history, etc., and delivers accurate, timely, and contextually grounded guidance. The contextual information we incorporate include the task description (comprising common task steps and approximate duration), the current task step, hand action recognition data, and conversation history between the user and the instructor. The task description is usually sourced from a knowledge base or manual, the current task step is identified by step detection models applied to the data captured by AR/VR-capable device, hand action data are derived from hand activity recognition models also applied to device-captured data, and dialogue history is collected from audio-capturing sensors. We do not delve into selecting or optimizing step detection and hand action recognition models, as these aspects are considered outside the scope of this work.

\begin{figure*}
    \centering
\includegraphics[width=1\textwidth]{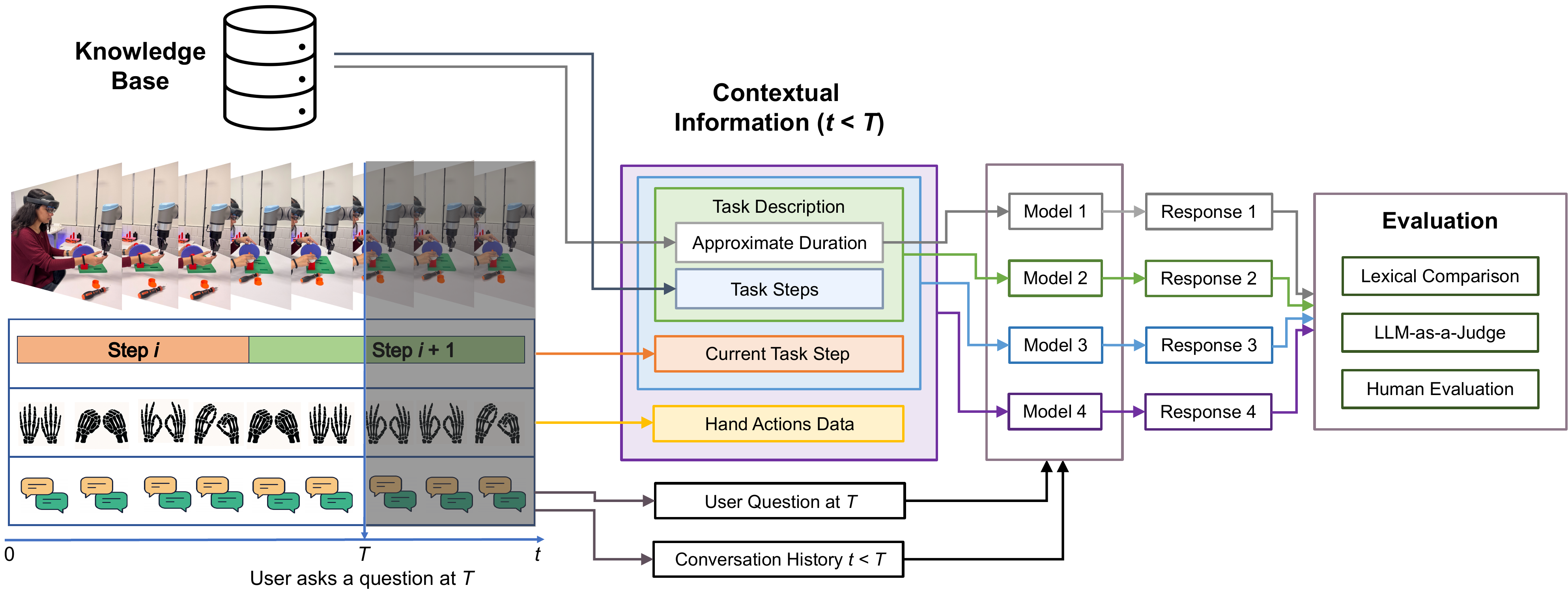}
    \caption{Illustration of our incremental prompting methodology. Contextual datasets, restricted to real-time information up to time $T$, are progressively provided to multiple models. Each model receives increasingly detailed context, starting from basic task duration information to detailed hand actions data. The models generate responses based on this incremental context, and these outputs undergo a multi-faceted evaluation, including lexical comparison, LLM-as-a-judge, and human evaluation.}
    \label{fig:flowchart}
\end{figure*}

As illustrated in Figure~\ref{fig:flowchart}, we consider multiple models in our incremental prompting strategy, each receiving progressively more detailed contextual information compared to the previous one. Importantly, all models receive the user's question at the current time $T$, and all contextual information provided to the model is restricted to data available up to this point (time $T$), ensuring that responses are generated based on real-time information without any knowledge of future events. This approach closely mimics real-world scenarios where a real-time assistant would not have access to future events or user actions. By limiting the input data strictly to historical and current context, we create realistic conditions for evaluating the real-time performance of the LLM assistant and its ability to provide meaningful assistance under practical constraints. 

\subsection{Models}
\label{subsec: models}
To systematically evaluate the effect of contextual richness on model performance, we designed four incremental model configurations, each incorporating progressively more detailed multimodal inputs. The following summarizes their configurations and intended purposes (see Figure~\ref{fig:flowchart}). All models were implemented using the Azure OpenAI API (\texttt{GPT-4o}, \texttt{version: 2025-01-01-preview}). 
Each query was executed with a maximum token limit of~100 to mimic concise, instructor-style responses. 
For detailed prompt templates of each model configuration, please refer to Appendix~\ref{appendix:prompts}.

    \subsubsection{Model 1—Minimal Context (Baseline)} This model receives only high-level task metadata (e.g., task type, duration, or goal). The input excludes any procedural or temporal context, simulating scenarios where the assistant has limited background information. 
    
\subsubsection{Model 2—Task Overview Context} This configuration adds structured task-step information, representing the full sequence of operations required to complete the task. This gives the model a procedural overview, enabling step-aware reasoning. \textit{Additional inputs:} structured step list (e.g., Step $1–N$ with brief descriptions).
    
\subsubsection{Model 3—Real-Time Step Context} Building upon Model 2, this version integrates temporal awareness by identifying the user’s current task step at the time a query is made (time $T$). This allows the model to tailor its response specifically to the user’s in-progress activity. \textit{Additional inputs:} current step index or label, task progress indicator, or timestamp alignment. 
    
\subsubsection{Model 4—Full Multimodal Context} The final configuration extends the context further by incorporating hand action data, encoded as verb–noun pairs (e.g., “pick tool,” “measure part”) extracted from real-time motion segmentation. These actions, typically spanning 1–2~seconds, provide fine-grained behavioral cues that enrich situational understanding. \textit{Additional inputs:} hand action segments, action-confidence scores, and optional dialogue snippets.

While the current models operate in a reactive manner—responding to user queries as they arise—future work could extend this framework toward proactive assistance, where the LLM anticipates user needs and autonomously offers timely clarifications or hints, further bridging the gap between automated systems and human instructors.


\subsection{Evaluation Metrics}
In this section, we elaborate on the mechanisms used to assess the proposed LLM assistant real-time performance and effectiveness, ensuring that the model's responses align closely with the user's immediate informational needs within the AR/VR-assisted task execution.
To evaluate the performance of real-time context-aware LLM assistant, we adopt three complementary approaches: lexical comparison, LLM-as-a-judge, and human evaluation.

\subsubsection{Lexical Comparison}
We begin with lexical comparison, employing both BERTScore \cite{Zhang2020} and cosine similarity between sentence embeddings computed via SentenceTransformer models \cite{Reimers2019}. These metrics measure the semantic similarity between the LLM assistant response and the ground truth answer. BERTScore leverages contextualized token embeddings to better capture paraphrastic overlap than traditional $n$-gram-based metrics such as BLEU \cite{Papineni2002} and ROUGE \cite{Lin2004, Schluter2017}. Similarly, cosine similarity between dense sentence embeddings captures broader semantic alignment, particularly useful when surface forms differ but the conveyed meaning remains intact. However, lexical metrics, even those based on contextual embeddings, struggle to fully evaluate the correctness, factual grounding, and coherence of longer, open-ended responses. They may assign high similarity to vague or irrelevant responses, or penalize correct answers phrased differently from the reference. Moreover, these metrics are inherently rely on the availability of a high-quality reference response, making them unsuitable for evaluating tasks where there is no single ground truth or multiple valid answers are possible.

\subsubsection{LLM-as-a-judge}
To overcome lexical comparison limitations, we adopt the increasingly popular paradigm of using a LLM as a judge. This approach, widely used in recent benchmarks and model evaluations \cite{Gu2025, Fu2023, Chu2024, Chiang2023, Wang2023n}, leverages an external LLM (e.g., GPT-4) to assess the quality of generated responses across multiple criteria such as correctness, completeness, relevance, and clarity. As noted by \cite{Gu2025}, LLM-as-a-judge frameworks have been applied across various domains, including text evaluation, model comparison, and decision-support tasks. Moreover, they offer a cost-effective and consistent alternative to large-scale human evaluations, reducing variability caused by subjective human biases. 
Despite their advantages, LLM-based evaluators also come with inherent limitations. Recent studies have identified issues such as positional bias, where the model tends to favor the first or longer response, as well as verbosity bias, self-consistency drift, and sensitivity to prompt phrasing \cite{Wang2023n, Fu2023, Wang2024fair}. Moreover, because the judging model often shares similar pretraining data and reasoning patterns with the evaluated models, it may reinforce shared misconceptions or stylistic artifacts, leading to overestimation of response quality. These factors underscore the importance of human validation and careful prompt design when interpreting LLM-as-a-judge results. To mitigate positional bias in our evaluation, we consider the average judgment scores across multiple permutations of model responses order, ensuring that no model is systematically advantaged by its position in the comparison prompt.

\subsubsection{Human Evaluation}
To complement automated evaluation, we conduct a human evaluation with expert annotators. Raters score each LLM assistant response on a 0–5 scale based on correctness, completeness, relevance, and clarity. While human evaluation remains the gold standard for natural language assessment, it also faces well-documented limitations, such as subjectivity, inter-rater variability, high cost, and limited scalability \cite{Chiang2023}. By combining one-time human validation with the proposed LLM-as-a-judge, our framework achieves both credibility and efficiency, paving the way for reproducible, cost-effective evaluation in future multimodal LLM systems.

\section{Experiments}
\label{sec: experiment}

In this section, we outline the experimental setup designed to evaluate our incremental prompting framework for real-time context-aware LLM assistants. Our primary goal is to examine how different contextual data modalities influence the LLM assistant ability to generate helpful and accurate responses. This analysis not only reveals how much contextual information is necessary to achieve reliable performance but also provides practical insight into how to select among data modalities under resource or bandwidth constraints. We describe the dataset and preprocessing steps, the detailed experimental procedure for prompting each model with varying levels of contextual input, and the evaluation methods employed. Together, these components form a systematic setup that enables us to investigate the effect of context granularity on real-time LLM assistants performance across realistic AR/VR task scenarios.

\subsection{HoloAssist Dataset}
The HoloAssist dataset \cite{Wang2023}, released by Microsoft Research, serves as a valuable benchmark for developing and evaluating context-aware assistants in AR environments. It contains rich, multimodal recordings of users performing real-world procedural tasks while being guided by human instructors. Captured using AR headsets, the dataset includes synchronized data streams such as egocentric RGB video, depth information, audio dialogues, hand pose and skeletal tracking, eye gaze, and detailed annotations. These diverse modalities enable the modeling of task progression, user intent, and context for real-time interactions. The dataset includes a variety of instructional scenarios, such as making coffee with a nespresso or espresso machine, assembling a computer, setting up printer components, fixing a motorcycle, and changing a circuit breaker. In this work, we utilize the HoloAssist dataset to simulate realistic AR task scenarios, providing the necessary contextual inputs for our incremental prompting framework and enabling comprehensive evaluation of the real-time context-aware LLM assistant performance under varying levels of context granularity.

In more detail, we consider sixteen types of tasks, each of which contains more than two user questions, making them strong candidates for evaluating the LLM assistant performance. Overall, we work with 141 task samples and a total of 625 questions. Ground truth responses are available for each question, allowing for quantitative comparison and evaluation of context-aware LLMs-generated responses across varying levels of contextual awareness. To effectively leverage the multimodal data provided by HoloAssist, we extract and process a set of contextual components that represent different levels of task information granularity used in our incremental prompting framework.

\subsection{Contextual Components}
\subsubsection{Task Description}
The task description represents the most fundamental layer of contextual information available while a user performs a task. It outlines the general sequence of steps typically followed from start to completion, along with an approximate estimate of the time required to complete the task. Although the HoloAssist dataset does not include a dedicated annotation for task descriptions, it has Narration annotation with a “Long-form Description” attribute. In this field, annotators were instructed to watch the videos and write descriptive paragraphs summarizing the observed activities.
To construct a unified task description for each task type, we leveraged the Narration annotations across multiple video samples. Specifically, we employed GPT-4 to aggregate and synthesize the common procedural steps shared among different instances of the same task. This process produced concise descriptions that capture both the overall workflow and the estimated duration of the task. The prompt used to extract task descriptions and approximate durations is provided in Appendix~\ref{appendix:task_description_prompt}.

\subsubsection{Current Task Step}
The current task step refers to the specific stage of the task that the user is performing at the moment a question is asked. It provides finer-grained contextual information compared to the overall task description, offering insight into the user’s immediate progress within the procedure. We utilize the Coarse-grained Action annotation in the HoloAssist dataset to represent this component. Each coarse action describes a high-level step in the task (e.g., change the battery of a GoPro in the GoPro setup task) and is associated with a start and end timestamp. When a user question occurs at time T, we identify the task step whose time interval encompasses T. If no such interval is found, we assign the placeholder phrase “The current step of the task is not specified. The student may not have started yet, or the step detection model failed to identify it.”

\subsubsection{Hand Action Data}
Hand action data represents the most detailed layer of contextual information available to the model, capturing fine-grained physical interactions from the beginning of the task up to the moment a question is asked. In the HoloAssist dataset, this information is provided under the Fine-grained Action annotations, which describe low-level atomic actions (e.g., press button, grab screw) in the form of verb–noun pairs. These annotations offer precise insights into what the user is doing and which objects they are interacting with at any given time, enabling the model to better interpret the user’s intent and generate contextually appropriate responses.

\subsubsection{Conversation History}
During task execution, all spoken interactions are captured through the audio sensors embedded in the AR headset. Since instructors guide task performers verbally, the HoloAssist dataset includes annotated dialogues that document these instructional exchanges. We use Conversation annotations to identify the task performer’s questions and the instructor’s corresponding answers, where instructor responses are specifically labeled under the Intervention Type: Answering Questions.

To ensure high-quality and complete conversational data, we instructed annotators to immerse themselves in each sample as if observing it live. This allowed them to accurately identify moments where the instructor provided hints or clarifications to the user. Annotators reviewed and refined instructor responses, editing incomplete or ambiguous answers when necessary, to produce a more reliable dataset for subsequent analysis.

The conversation history serves as a contextual component representing prior dialogue between the task performer and the instructor. For each question, we track preceding question–answer pairs and provide them as input to the models. This allows the system to maintain conversational continuity and generate responses that are coherent with the prior dialogue context.

\begin{table*}[ht]
\centering
\caption{Summary of contextual inputs for each model in the incremental prompting framework.}
\label{tab:models}
\begin{tabular}{lccccc}
\toprule
\textbf{Model} & \multicolumn{2}{c}{\textbf{Task Description}} & \textbf{Current Task Step} & \textbf{Hand Action Data} & \textbf{Conversation History} \\
\cmidrule(lr){2-3}
 & \textbf{Approx. Duration} & \textbf{Task Steps} & & &  \\
\midrule
Model 1 & \checkmark & & & &\checkmark \\
Model 2 & \checkmark & \checkmark & & &\checkmark \\
Model 3 & \checkmark & \checkmark & \checkmark & & \checkmark \\
Model 4 & \checkmark & \checkmark & \checkmark & \checkmark & \checkmark \\
\bottomrule
\end{tabular}
\end{table*}

\subsection{Models}
As detailed in Section~\ref{subsec: models} and illustrated in Figure~\ref{fig:flowchart}, our incremental prompting framework comprises multiple models, each receiving progressively richer contextual information. All models take the user’s question at time $T$ as input, with contextual data limited to what is available up to that point, ensuring consistent real-time reasoning without access to future events. This setup closely mirrors realistic interactive conditions for AR/VR assistants. Table~\ref{tab:models} summarizes each model’s configuration and the corresponding contextual components used in our experiments.

\subsection{Prompt Design and API Configuration}
All models were implemented using the Azure OpenAI API through a Python interface, following a standardized configuration to ensure consistent evaluation across experiments. Each API call contained structured messages composed of a system instruction and a user query. To prioritize accuracy and consistency over creativity, the temperature was set to 0.5. This configuration maintains stable responses with limited randomness, reducing stylistic variability while preserving contextual precision. Since instructor responses in the dataset are typically short and concise, the maximum token length was limited to 100. Each prompt consisted of three components: (1) a system instruction defining the model’s role as a task guidance assistant, and (2) a user input including the user’s question along with (3) all contextual information available up to the current time step ($T$). Higher-level models received progressively richer context, as summarized in Table~\ref{tab:models}. Full prompt templates for all four models are provided in Appendix~\ref{appendix:prompts}. All model outputs were programmatically logged and subsequently evaluated using a combination of lexical comparison, LLM-as-a-judge, and human evaluation to comprehensively assess responses quality.

\section{Results}
\label{results}

This section compares the four models using lexical metrics, LLM-as-a-judge scoring, and human evaluations, and further analyzes LLM–human agreement and the framework’s sensitivity to individual elements through ablation studies.

\subsection{Lexical Comparison}
We computed lexical and embedding-based similarity metrics for the responses generated by each model. Specifically, we report BERTScore precision, recall, and F1 \cite{Zhang2020}, along with cosine-based semantic similarity using SentenceTransformer embeddings \cite{Reimers2019}, for all four models (Table~\ref{tab:quantitative}). These metrics quantify the degree of overlap and semantic alignment between model-generated responses and the reference instructor answers.
While the results show minor numerical variations across models, the overall trend does not yield a clear or consistent conclusion regarding which contextual configuration performs best. The scores remain relatively close, suggesting that these lexical metrics may not be sufficiently sensitive to capture nuanced improvements arising from richer multimodal context. Moreover, both BERTScore and sentence-level cosine similarity depend heavily on lexical overlap and reference phrasing, making them less reliable for evaluating open-ended, context-grounded responses. They may overrate vague or partially accurate answers while underrating responses that are semantically correct but expressed in different wording.
These observations highlight the limitations of using automated similarity-based metrics alone for assessing real-time AR/VR assistants. Therefore, we complement these quantitative results with LLM-based and human evaluations to provide a more comprehensive understanding of models behavior and contextual reasoning quality.

\begin{table}[!t]
\centering
\caption{Quantitative results using BERTScore and Semantic Similarity metrics.}
\label{tab:quantitative}
\begin{tabular}{lcccc}
\hline
\textbf{Model} & \multicolumn{3}{c}{\textbf{BERTScore}} & \textbf{Semantic Sim.} \\ 
\cline{2-4}
 & \textbf{F1} & \textbf{Precision} & \textbf{Recall} &  \\ \hline
Model 1 & 0.54 & 0.47 & 0.65 & 0.18 \\
Model 2 & 0.53 & 0.46 & 0.65 & 0.19 \\
Model 3 & 0.53 & 0.46 & 0.65 & 0.18 \\
Model 4 & 0.51 & 0.44 & 0.64 & 0.18 \\ \hline
\end{tabular}
\end{table}

\subsection{LLM-as-a-Judge}
\label{sec:llmjudge}
To complement the quantitative metrics, we employed a LLM as an automatic judge to assess response quality more holistically. The judge model was provided with all contextual information available up to the point each question was asked, including task description, current task step, hand actions data, and conversation history, and was instructed to evaluate the four model-generated responses based on \textit{correctness}, \textit{completeness}, \textit{contextual relevance}, and \textit{clarity}. This LLM-based evaluation enables the capture of semantic and pragmatic aspects that lexical similarity metrics cannot reliably measure.
To mitigate potential positional bias in the judging process (i.e., the tendency of an LLM to favor responses appearing earlier in the list), we randomized the order of the four model outputs across four different permutations: 1234, 3421, 4321, and 2143, where each number represents the index of a model in the evaluation prompt (e.g., in permutation 3421, Model~3 is presented first, followed by Models~4, 2, and 1). While exhaustively evaluating all $4!$ orderings would be computationally expensive, we selected four representative permutations that allow each model to appear in every possible position, which we believe sufficiently mitigates positional bias. The final score for each model represents the average across these permutations. 
Each model received individual 0–5 ratings on all dimensions, along with an overall comparison summary identifying the most contextually appropriate response. The full judging prompt template is provided in Appendix~\ref{appendix:judge_prompt}. 

\begin{figure}[!t]
    \centering
    \includegraphics[width=1\linewidth]{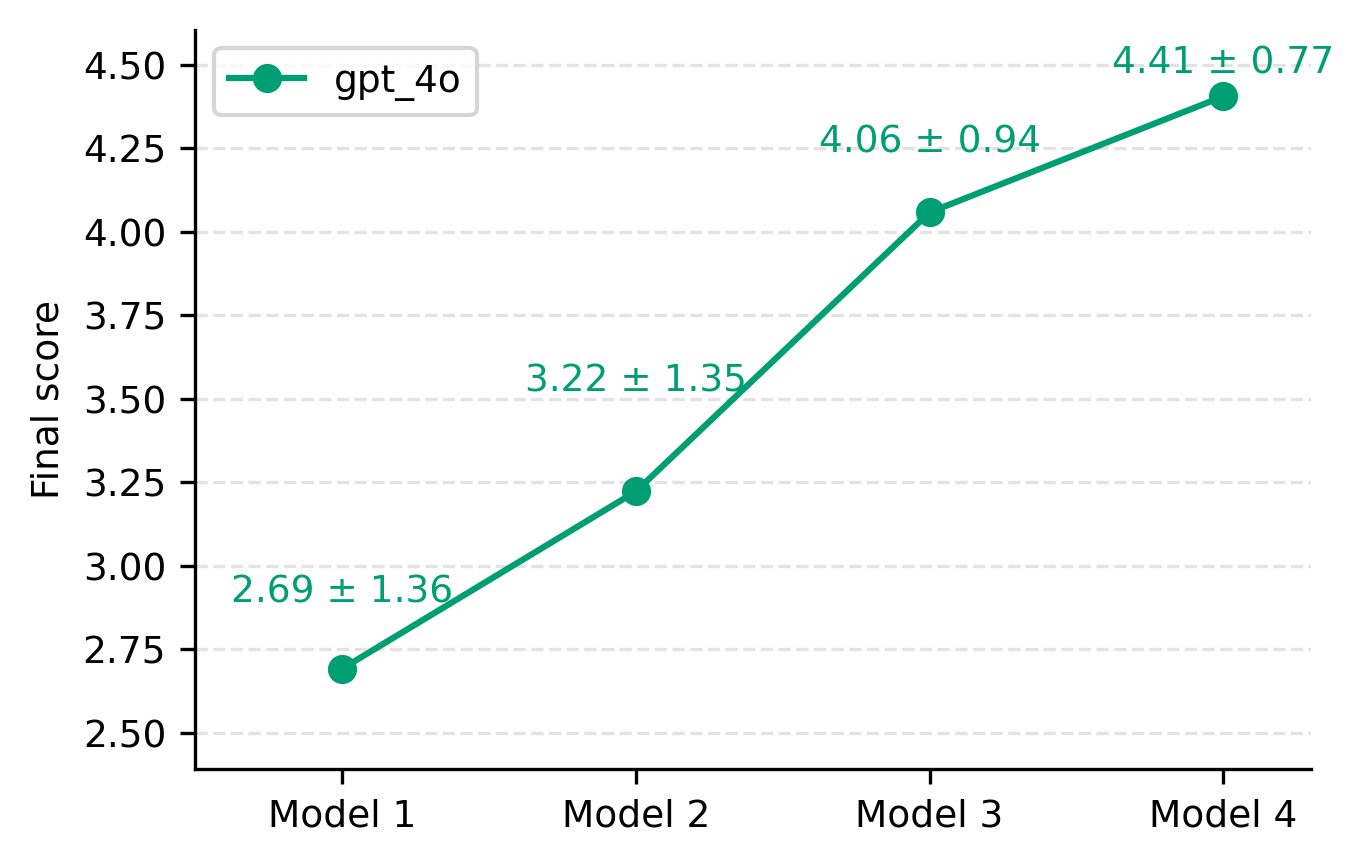}
    \caption{Average final score (mean ± standard deviation) across four models as evaluated by the proposed LLM-as-a-judge. Performance improves consistently as additional contextual signals are introduced, with Model 4 achieving the highest accuracy and consistency.}
    \label{fig:llm_finalscore}
\end{figure}

As shown in Figure~\ref{fig:llm_finalscore}, performance improves consistently across models as more contextual information are introduced. The mean final score rises from 2.69 for Model 1 to 4.41 for Model 4, indicating that incremental addition of task-related information (e.g., current task step, hand actions) enables the model to generate more accurate and contextually grounded responses. Moreover, the standard deviation of scores decreases from $±1.36$ in Model~1 to $±0.77$ in Model~4, indicating that richer contextual inputs not only enhance performance but also lead to more consistent and reliable responses. 

\begin{figure}[!t]
    \centering
    \includegraphics[width=1\linewidth]{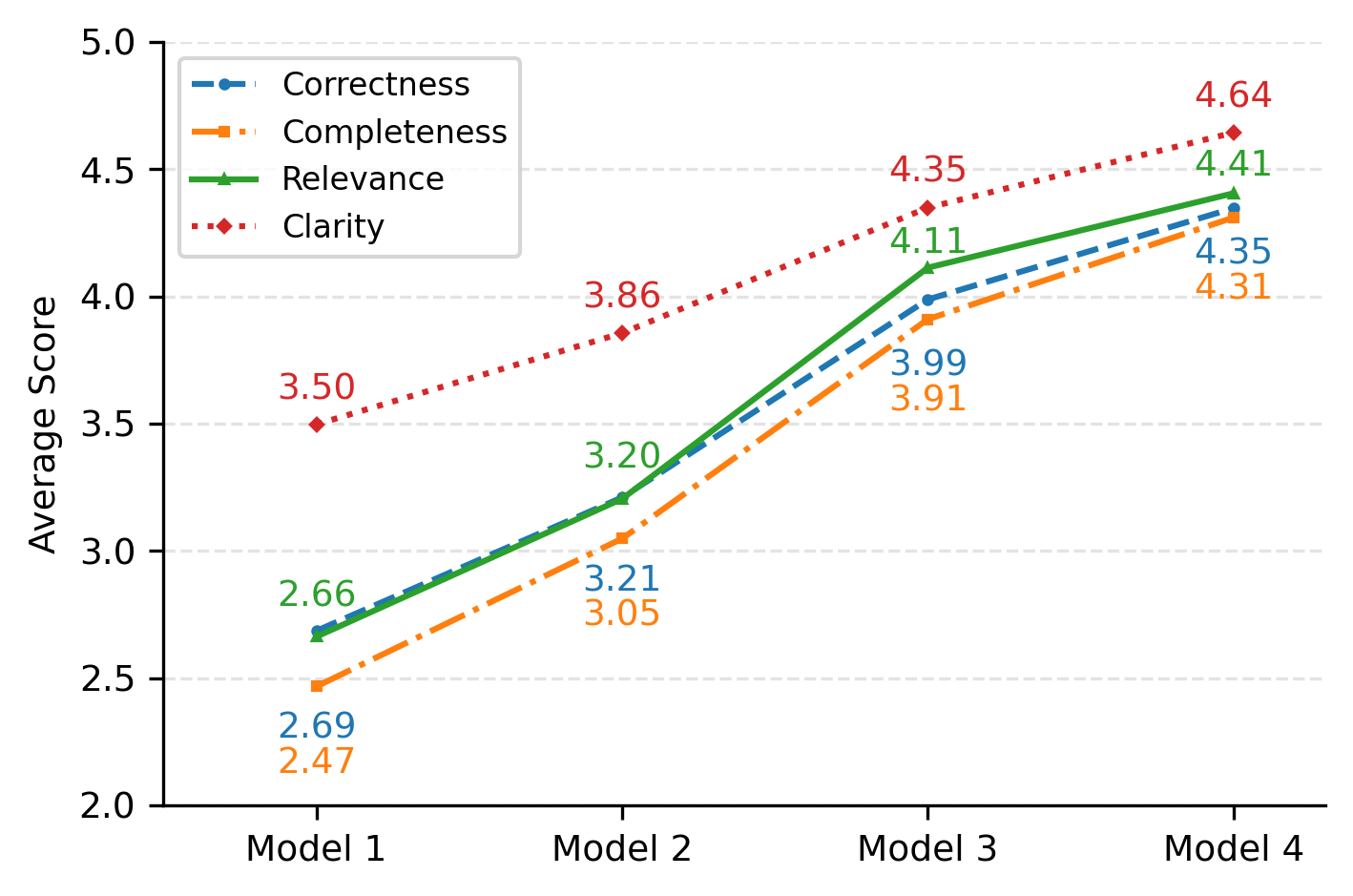}
    \caption{LLM-as-a-judge's dimension-wise evaluation scores for each model, showing trends across correctness, completeness, contextual relevance, and clarity. All dimensions exhibit consistent improvement as contextual components are added, with Model~4 achieving the highest average scores in all categories.}
    \label{fig:llm_dimensionwise}
\end{figure}

To further dissect the model’s performance, Figure~\ref{fig:llm_dimensionwise} presents the average scores across four evaluation dimensions, correctness, completeness, contextual relevance, and clarity. Each dimension follows a similar upward trend, reinforcing that richer contextual inputs improve not only overall accuracy but also the interpretability, coherence, and task alignment of generated responses.

\begin{figure}[!t]
        \centering
        \includegraphics[width=1\linewidth]{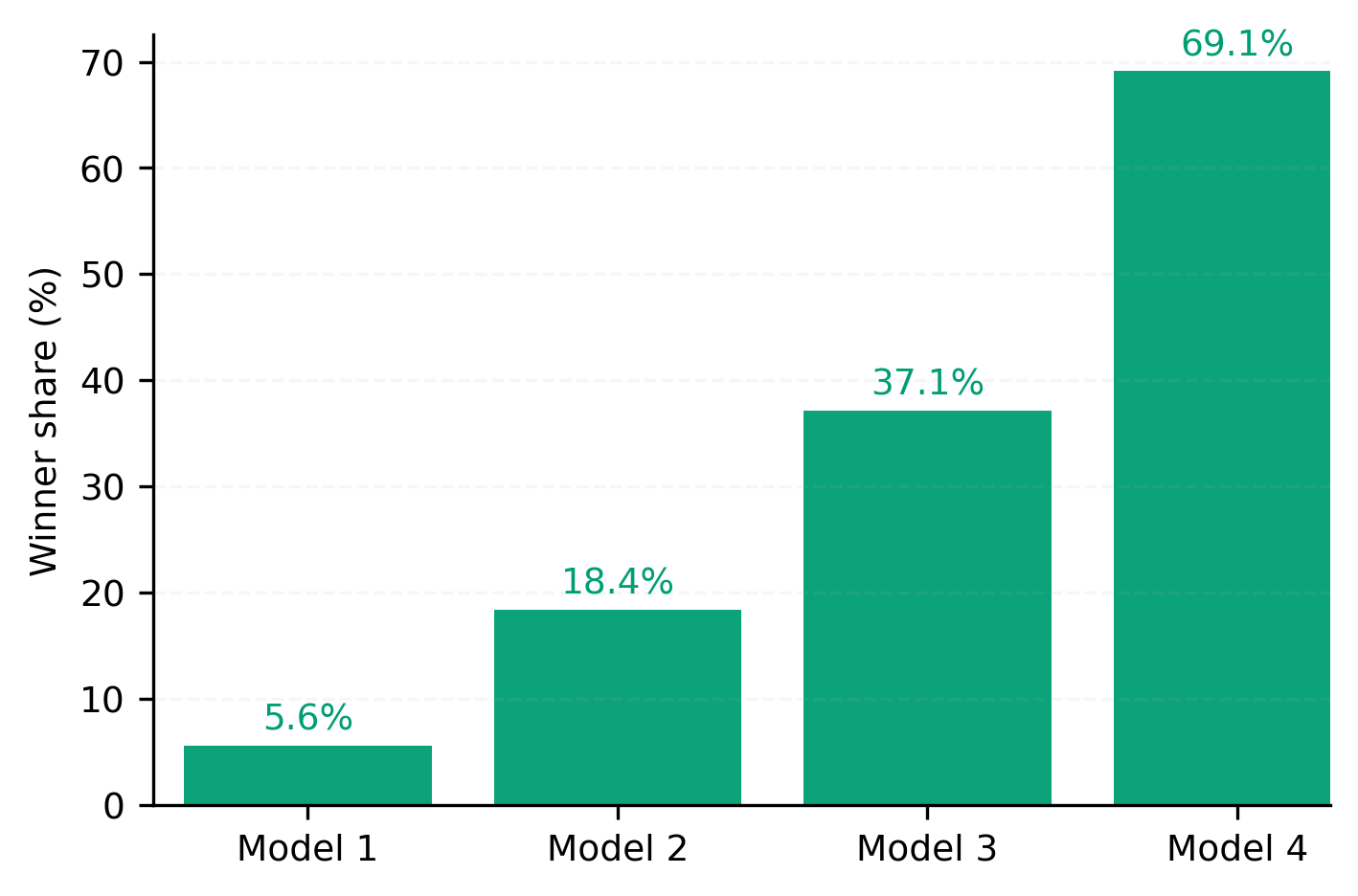}
        \caption{Winner share percentage across models based on LLM-as-a-judge evaluations. Model 4 dominates with $69.1\%$ of top-ranked responses, indicating that contextual enrichment strongly enhances perceived response quality.}
    \label{fig:llm_winnershare}
\end{figure}

Similarly, the winner share distribution in Figure~\ref{fig:llm_winnershare} exhibits a consistent upward trend, with Model~4 judged as the best-performing response in $69.1\%$ of cases, compared to only $5.6\%$ for Model~1. The winner for each question was determined based on the model receiving the highest final score from the LLM-as-a-judge; in cases where two or more models achieved the same top score, all tied models were counted as winners. Consequently, the total winner share exceeds $100\%$, reflecting overlapping wins in tied evaluations. This monotonic improvement highlights the strong positive effect of contextual enrichment on the model’s reasoning accuracy and procedural precision.  

\subsubsection{Robustness Across LLM Variants}
\begin{figure}[!t]
    \centering
    \includegraphics[width=1\linewidth]{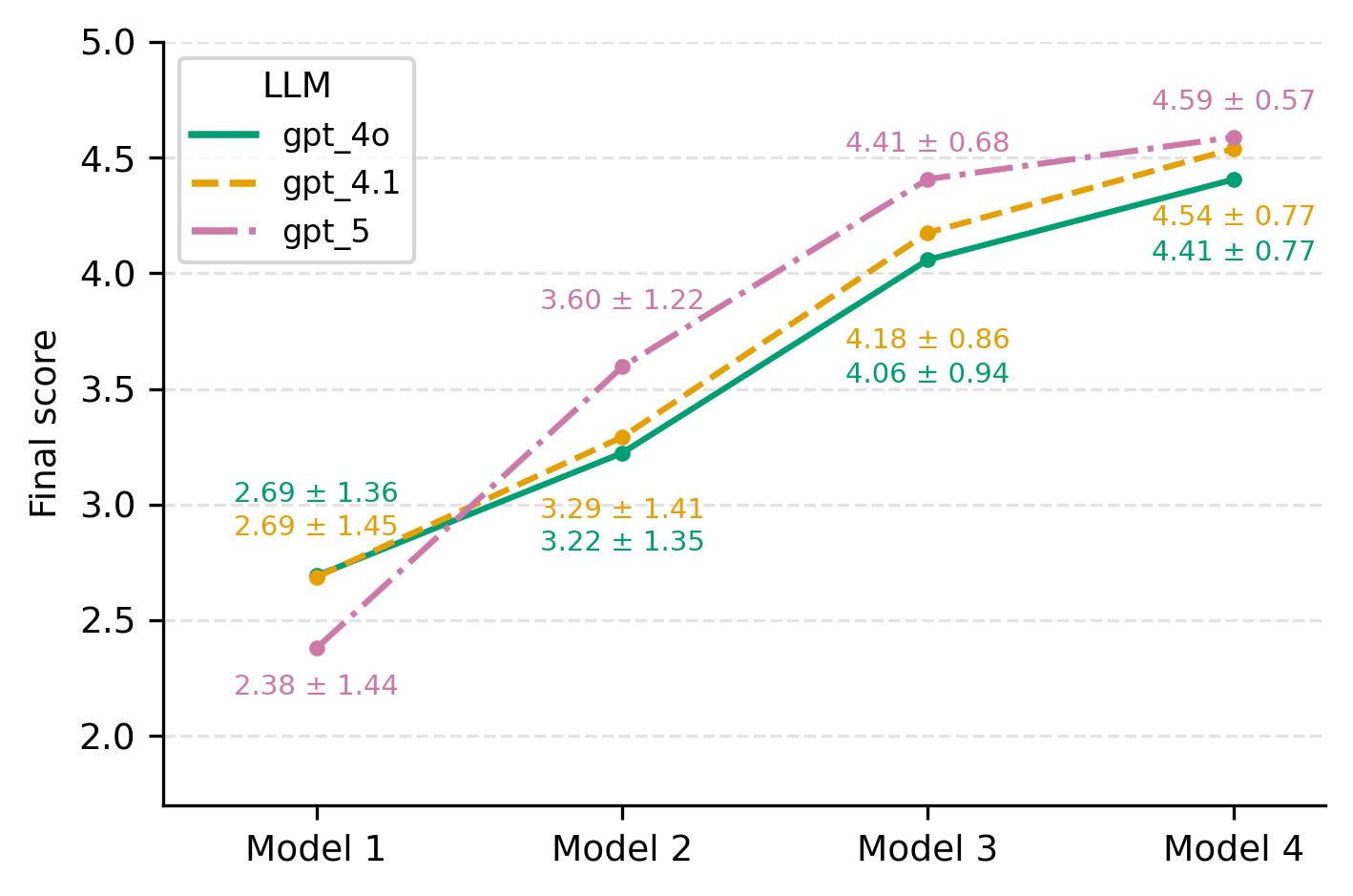}
    \caption{Robustness analysis of the \textit{LLM-as-a-judge} framework using different judge models (GPT-4o, GPT-4.1, and GPT-5). All exhibit a consistent upward trend, demonstrating that the framework’s evaluations are stable across different models.}
    \label{fig:llm_robustness}
\end{figure}

To evaluate the robustness and model-agnostic nature of our \textit{LLM-as-a-judge} framework, we replicated the evaluation using three different LLM models: GPT-4o (\texttt{version: 2025-01-01-preview}), GPT-4.1 (\texttt{version: 2025-01-01-preview}), and GPT-5 (\texttt{version: 2025-04-01-preview}). As illustrated in Figure~\ref{fig:llm_robustness}, all models exhibit the same monotonic improvement pattern across the four contextual configurations. The average final scores increase steadily from Model~1 to Model~4 for all judges, with only minor numerical differences in absolute scores.

This strong consistency across distinct LLM architectures underscores the reliability and reproducibility of our evaluation framework. Despite variations in model families and underlying reasoning capabilities, all judges converge on the same relative ranking of models, confirming that the observed performance gains are not tied to a specific evaluator. This model-agnostic stability reinforces the validity of our incremental prompting approach and highlights the potential of our LLM-as-a-judge framework as a dependable, scalable evaluation methodology.

\subsection{Human Evaluation}

We further conducted a human evaluation to assess the quality of model-generated responses from an expert perspective. Human annotators rated each response on a 0–5 scale across four dimensions: correctness, completeness, contextual relevance, and clarity. The same task-level contextual information was provided to the annotators to ensure fair and consistent judgment across models.

\begin{figure}
    \centering
    \includegraphics[width=1\linewidth]{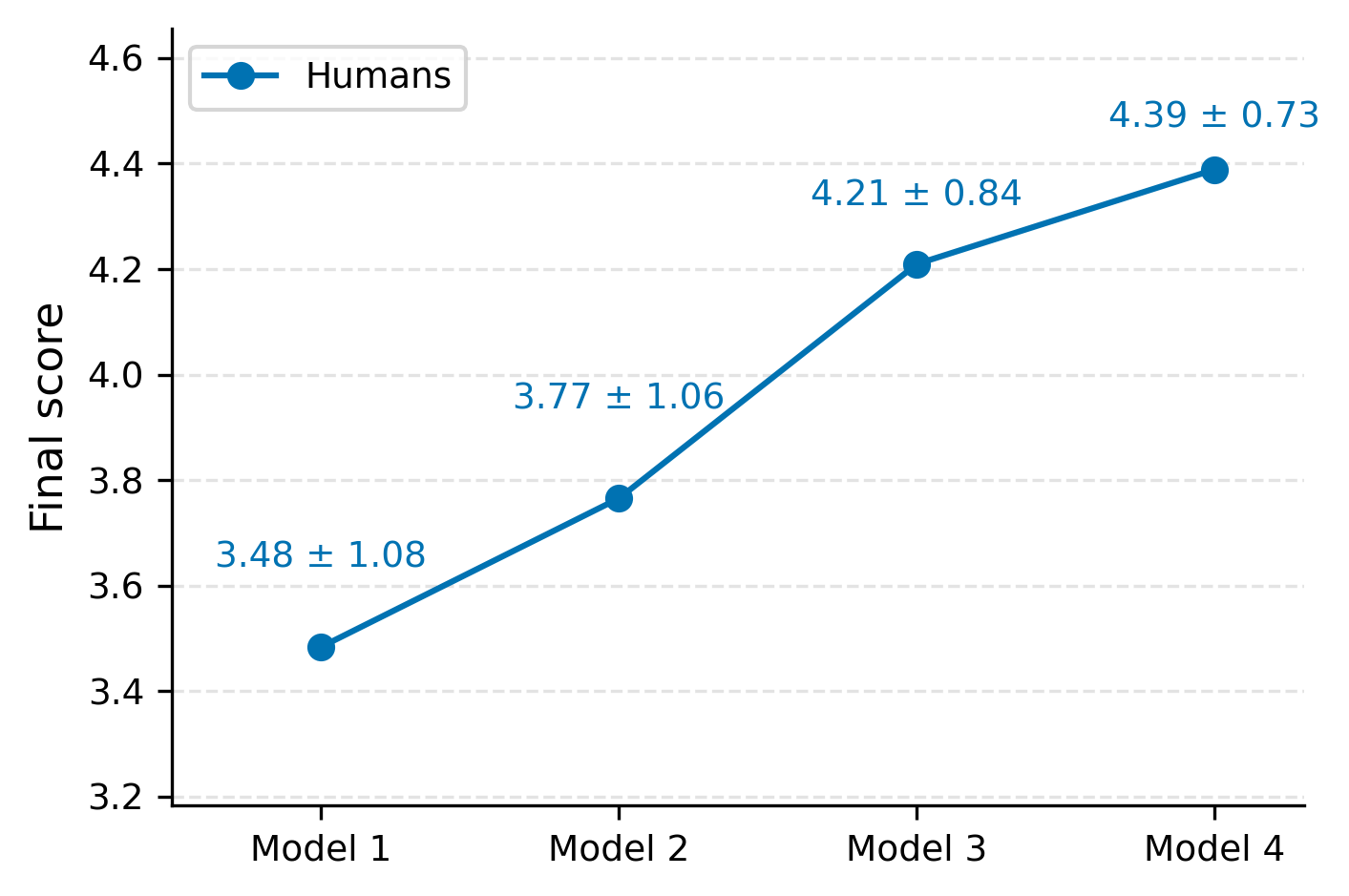}
    \caption{Human evaluation (mean ± standard deviation) results for all models. The mean final score increases consistently from Model~1 to Model~4, with lower variance at higher context levels, indicating improved response quality and stability.}
    \label{fig:human_finalscore}
\end{figure}

As shown in Figure~\ref{fig:human_finalscore}, the results exhibit a clear upward trajectory in response quality as contextual inputs increase. The mean final score improves steadily from $3.48 \pm 1.08$ in Model~1 to $4.39 \pm 0.73$ in Model~4, suggesting that the progressive inclusion of task descriptions, current task step, and hand actions data enables the models to produce more accurate and contextually grounded instructional responses. Moreover, the decreasing standard deviation across models indicates a rise in response consistency as richer information becomes available.

\begin{figure}
    \centering
    \includegraphics[width=1\linewidth]{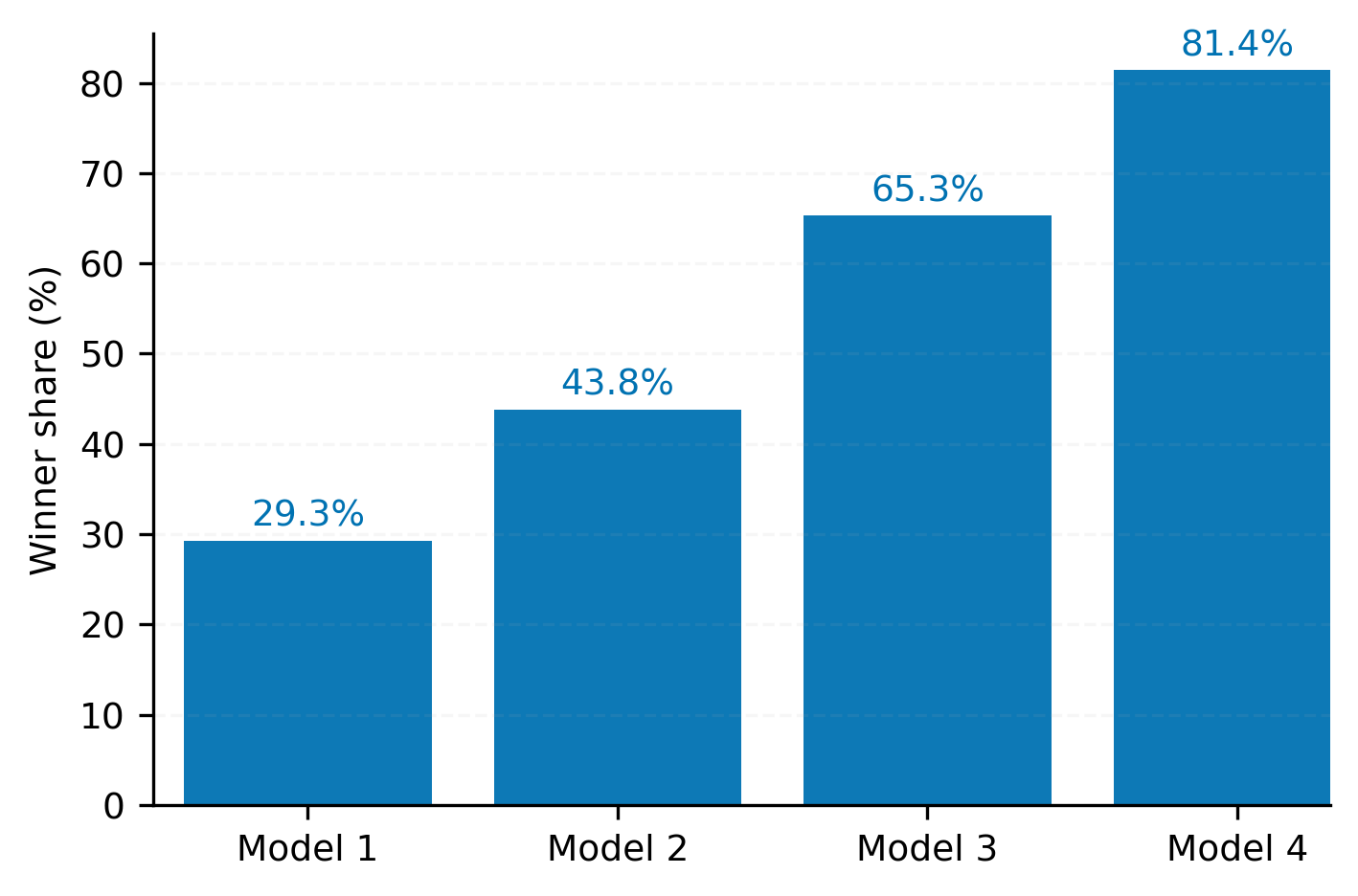}
    \caption{Winner share distribution from human evaluation. Each bar represents the proportion of instances where a model was rated highest by human evaluators. Model~4 dominates with winning $81.4\%$ of cases, demonstrating the advantage of richer contextual grounding. Percentages exceed $100\%$ due to ties across models.}
    \label{fig:human_winnershare}
\end{figure}

A similar trend is reflected in the winner share distribution, Figure~\ref{fig:human_winnershare}, where Model~4 was judged as the best-performing response in $81.4\%$ of cases, followed by Model~3 with $65.3\%$, Model~2 with $43.8\%$, and Model~1 with only $29.3\%$. The winner for each question was determined based on the model receiving the highest average human score; when multiple models tied for the top score, all were counted as winners, leading to cumulative percentages exceeding $100\%$.

Overall, these results highlight that human evaluators consistently favor models equipped with richer contextual understanding, reinforcing the strong positive effect of multimodal and temporal cues on the perceived quality, relevance, and precision of real-time instructional responses.

\subsection{LLM–Human Agreement}

\begin{figure}[!t]
    \centering
    \includegraphics[width=1\linewidth]{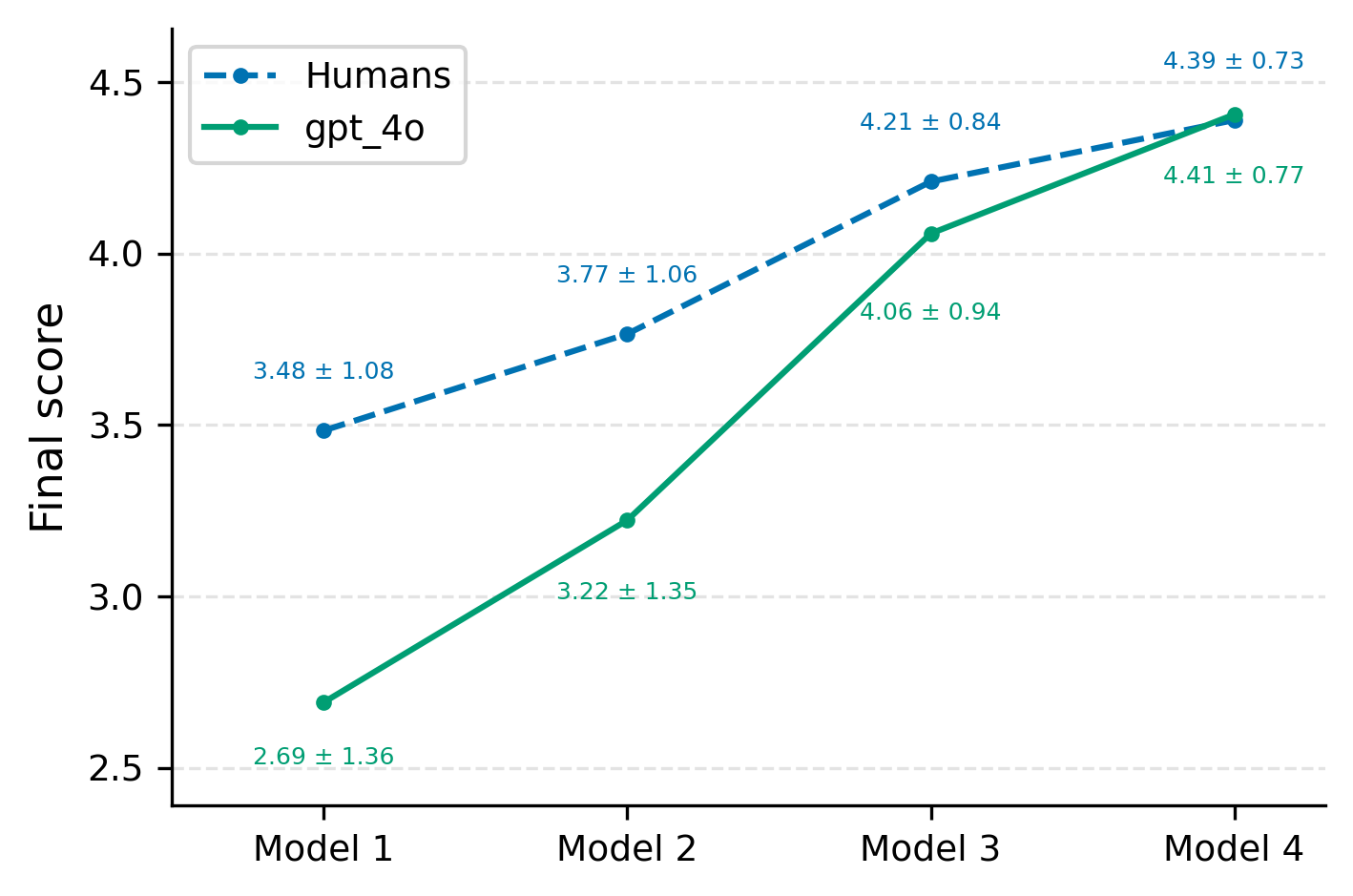}
    \caption{Comparison of LLM-as-a-judge and human evaluation trends. Both follow a similar improvement trajectory across models, confirming the high alignment between automated LLM-as-a-judge and human judgment.}
    \label{fig:llm_human_agreement}
\end{figure}

To assess the alignment between LLM-as-a-judge and human evaluations, we compared both evaluators’ judgments of models performance. As shown in Figure~\ref{fig:llm_human_agreement}, the trends in final scores assigned by humans and LLM-as-a-judge follow a monotonic pattern, with consistent improvement from Model~1 to Model~4. This similarity indicates that both evaluators perceive contextual enrichment, such as current task step and hand actions data, as key factors contributing to higher response quality and stability.

\begin{figure}[!t]
    \centering
    \includegraphics[width=1\linewidth]{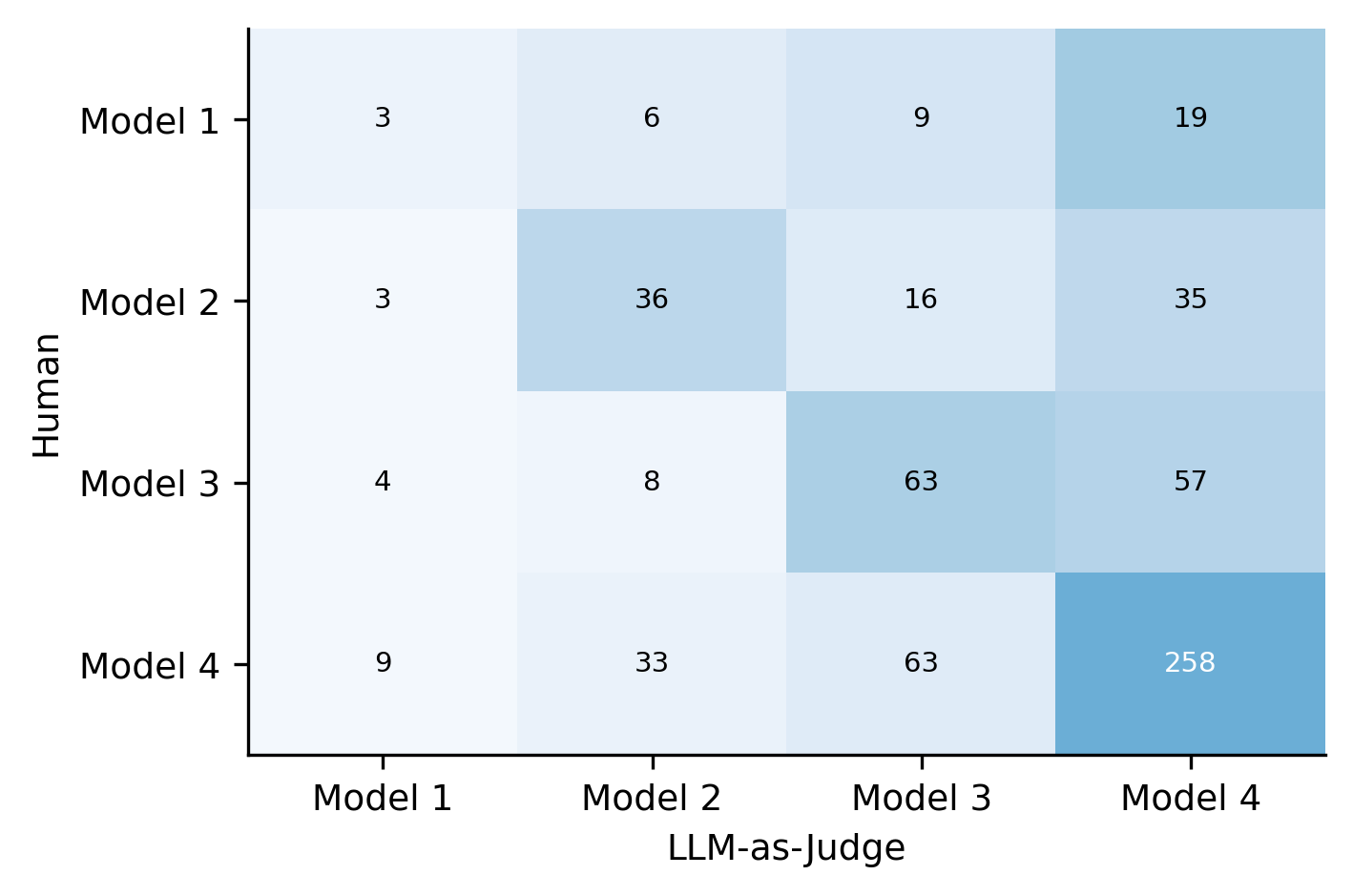}
    \caption{Agreement matrix between LLM and human winners. Each cell represents the number of shared wins for a given model pair. An overall agreement rate of $83.7\%$ indicates strong consistency between LLM-as-a-judge and human evaluators. Multiple winners per case were included to account for ties.}
    \label{fig:agreement_matrix}
\end{figure}

A more detailed comparison is illustrated in the agreement matrix in Figure~\ref{fig:agreement_matrix}, which visualizes the correspondence between human and LLM-as-a-judge identified winners. Since each instance could have multiple winners in cases of tied scores, all were counted toward agreement. The results show an overall $83.7\%$ agreement between human and LLM judgments, which is close to the agreement among
humans \cite{Zheng2023}, with the highest concordance occurring for Model~4, confirming that both evaluators consistently recognized its superior contextual reasoning. This pattern aligns with prior observations from \cite{Zheng2023}, where GPT-4 demonstrated stronger agreement with human evaluators when the performance gap between models was more pronounced, suggesting that LLM judges are particularly reliable in distinguishing clear quality differences.

Furthermore, the winner share distribution across both evaluators revealed closely aligned patterns, Figures~\ref{fig:llm_winnershare} and~\ref{fig:human_winnershare}, reinforcing that the LLM judge captured the same performance hierarchy as human evaluators. This strong consistency demonstrates that our LLM-as-a-judge framework can reliably emulate human-level evaluation of real-time, multimodal task assistance. Given that human evaluations are resource-intensive and difficult to scale, these findings highlight the potential of the proposed LLM-as-a-judge evaluation framework as a cost-effective and reproducible alternative for large-scale or continuous benchmarking of AR/VR instruction models.

\subsection{Ablation Studies}
To better understand the contribution of individual contextual components to the model’s performance, we conduct a two-part ablation study designed to reveal both their unique importance and interdependencies. This analysis also provides practical guidance for deploying the framework under varying resource or data-availability constraints. 

Specifically, we perform:
(\textit{i}) an All-Except-One (AEO) analysis, in which each context type is removed while keeping the others intact, allowing us to quantify the performance degradation caused by the absence of that component; and
(\textit{ii}) an Only-One (OO) analysis, in which the model receives only a single contextual input at a time, revealing which component alone contributes most to the performance when multimodal information is limited. 
In both analyses, model performance was assessed using the \textit{LLM-as-a-judge} framework described in Section~\ref{sec:llmjudge}, ensuring consistent and context-aware evaluation of response quality across different configurations. These complementary experiments reveal both the individual and combined effects of contextual modalities on the model’s ability to generate accurate, contextually grounded responses.

\subsubsection{All-Except-One (AEO)}

In the AEO setup, each model omits one contextual component while retaining the others to isolate its contribution to overall performance. Specifically, Model~I excludes the \textit{task duration}, Model~II omits the \textit{task steps}, Model~III lacks the \textit{current task step}, and Model~IV excludes the \textit{hand action data}. Removing any single component leads to a measurable decline compared to the full “All-in” configuration, confirming that each context type contributes meaningfully to the model’s overall comprehension and response generation.

\begin{figure}[!t]
    \centering
    \includegraphics[width=1\linewidth]{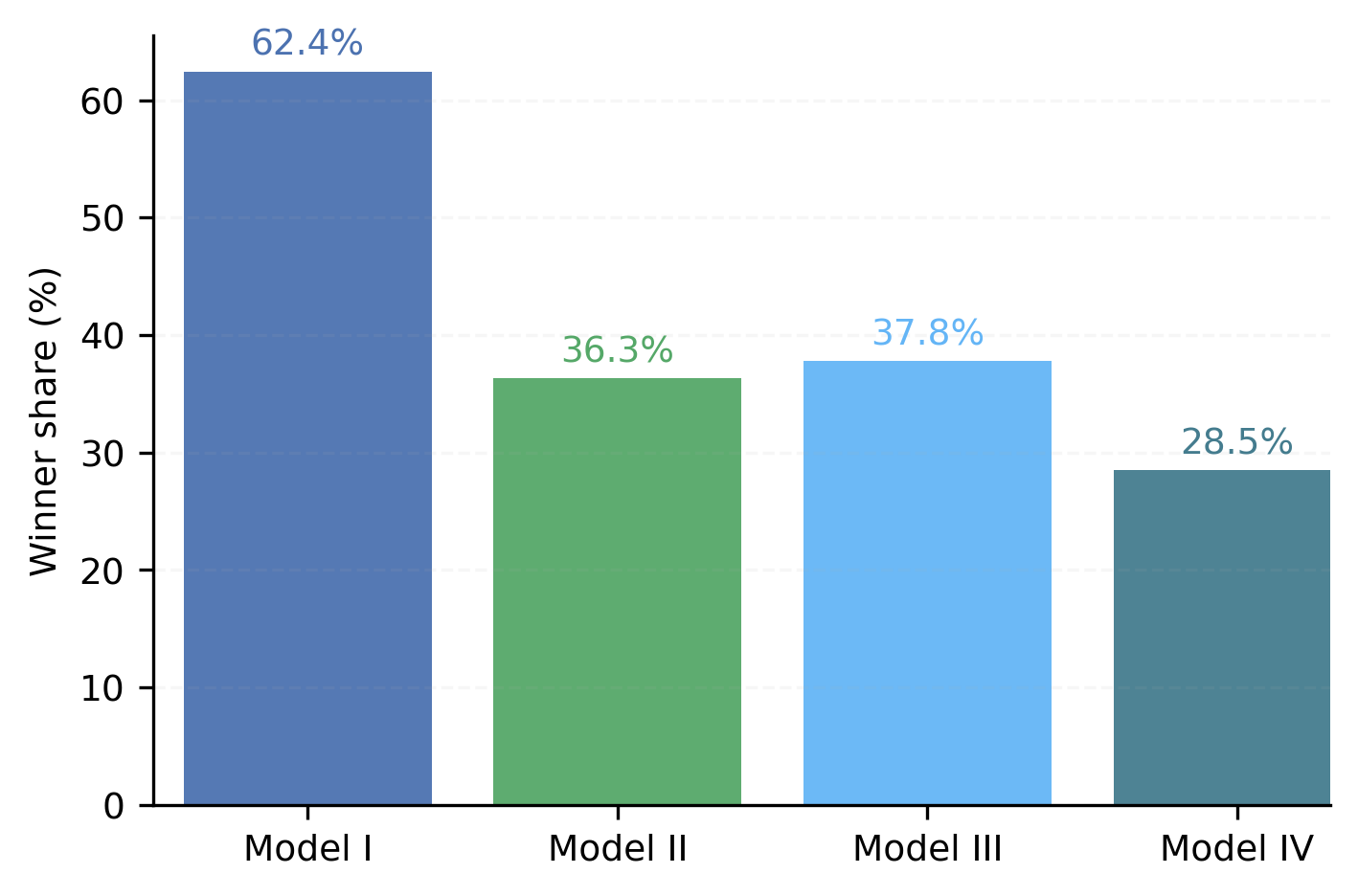}
    \caption{Winner share comparison across models in the All-Except-One (AEO) setup. The models missing task steps or current task step information (Models~II and~III) show similar declines. The absence of hand action data (Model~IV) results in the steepest drop, underscoring its critical role in maintaining robust model performance.}
    \label{fig:aeo_winner}
\end{figure}

\begin{figure}[!t]
    \centering
    \includegraphics[width=1\linewidth]{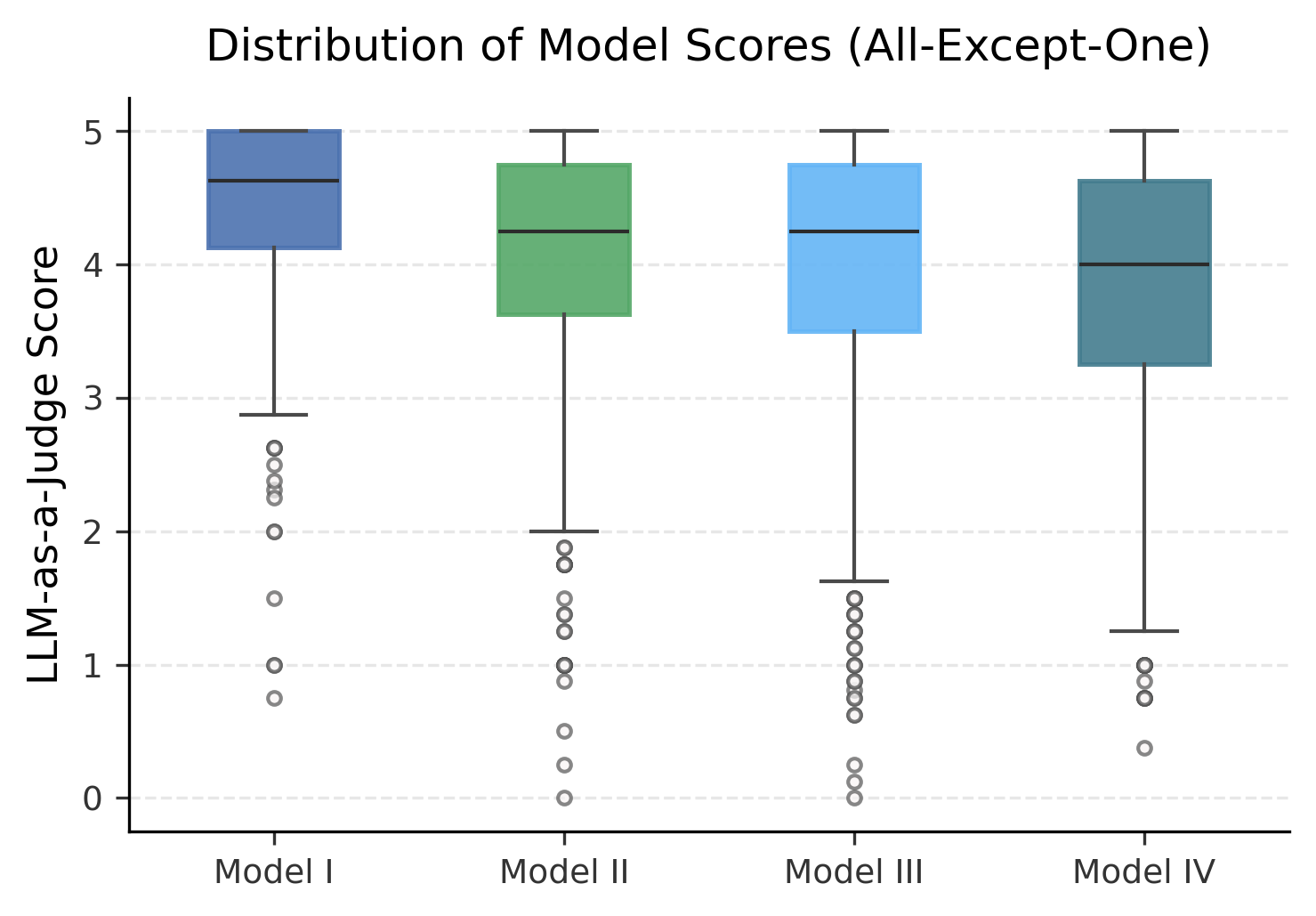}
    \caption{Boxplot of evaluation scores for the All-Except-One (AEO) analysis. Each variant omits one contextual component from the full “All-in” configuration. Removing any input source leads to a decline in performance, with the largest degradation observed when hand action data are excluded (Model~IV). The results demonstrate that all contextual components contribute meaningfully to generating accurate, contextually grounded responses.}
    \label{fig:aeo_boxplot}
\end{figure}

Figures~\ref{fig:aeo_winner} and~\ref{fig:aeo_boxplot} show that among the four variants, the absence of \textit{task description} (Model~II) and \textit{current task step} (Model~III) results in comparable degradation, indicating that both high-level and fine-grained task information play equally critical roles in grounding the model’s responses. Meanwhile, the exclusion of the \textit{hand action} modality (Model~IV) causes the most pronounced performance drop, suggesting that action-level cues provide uniquely valuable fine-grained context that other inputs cannot fully replace.
In general, the AEO results highlight the complementary and synergistic nature of contextual modalities: removing any component weakens the situational basis of the model, with the greatest sensitivity observed when hand action data are not available. These findings emphasize that integrating multiple context sources, particularly fine-grained action and step information, is essential for achieving robust and accurate real-time assistance.

\subsubsection{Only-One (OO)}

\begin{figure}
    \centering
    \includegraphics[width=1\linewidth]{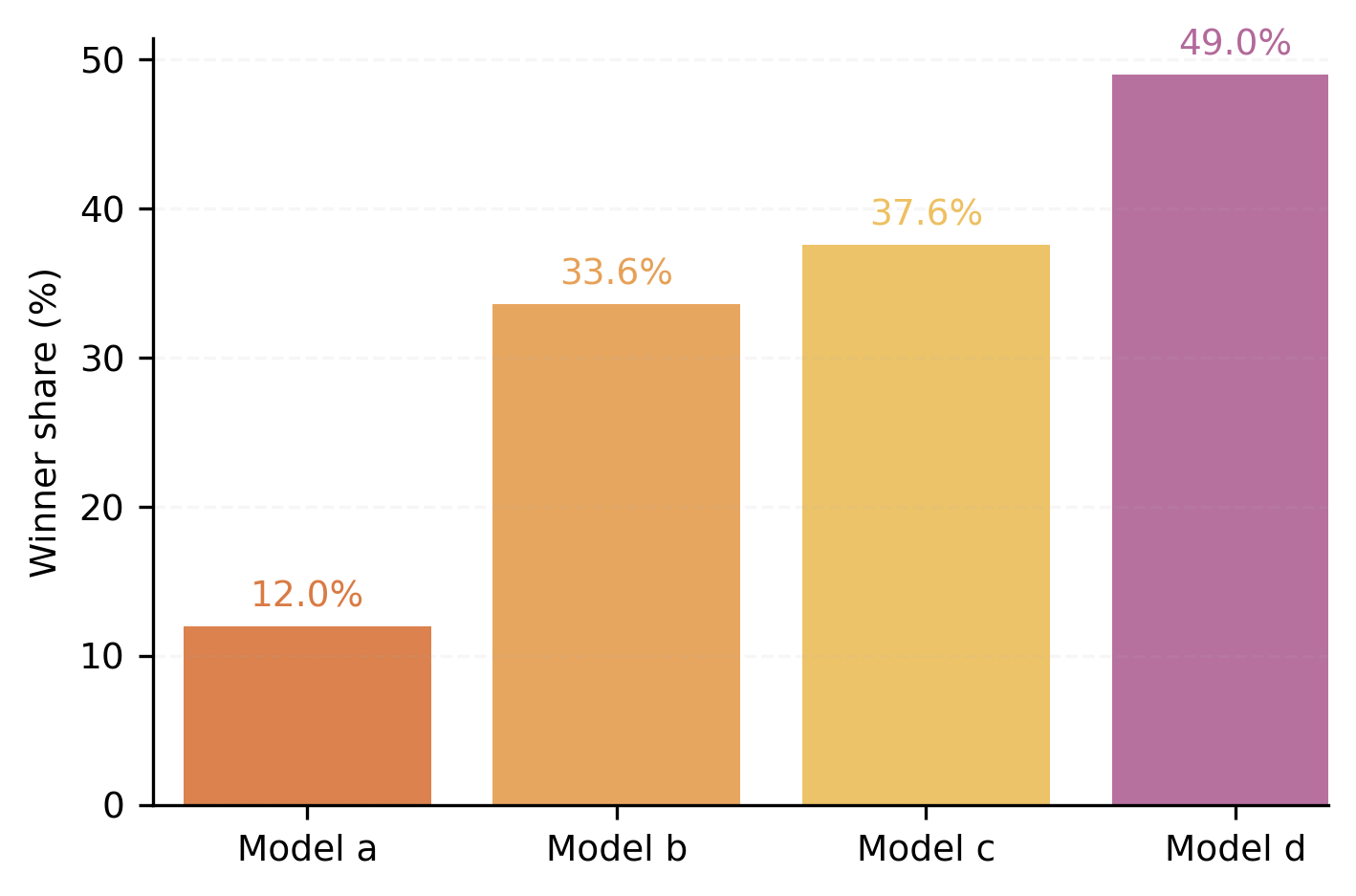}
    \caption{Winner-share comparison for the Only-One (OO) analysis. Each bar indicates the proportion of cases where a model’s response achieved the highest quality score according to LLM-as-a-Judge evaluation. The \textit{hand action} context (Model~d) attains the highest winner share, reinforcing its importance as the most reliable standalone signal for real-time task understanding.}
    \label{fig:oo_winner}
\end{figure}

\begin{figure}
    \centering
    \includegraphics[width=1\linewidth]{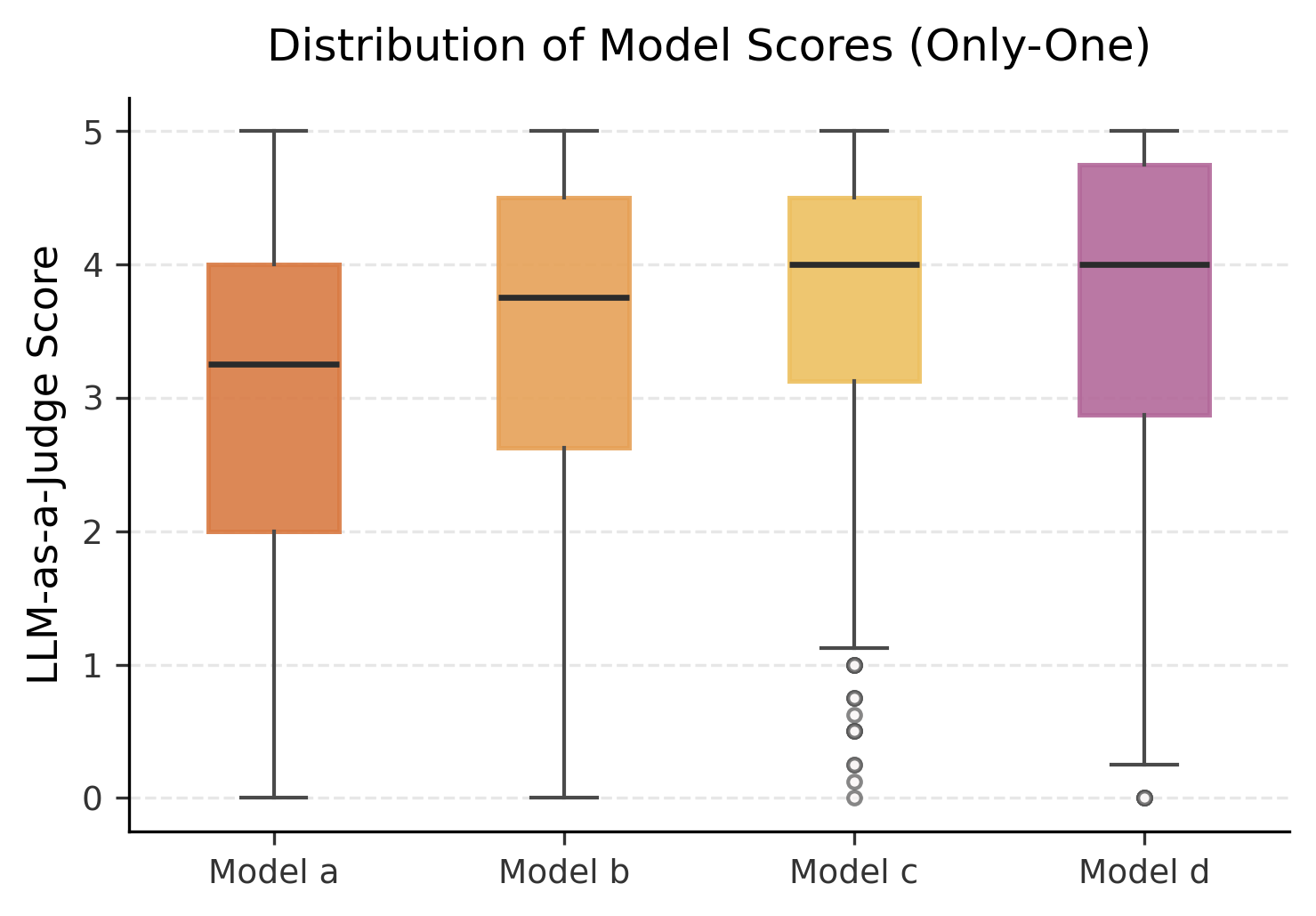}
    \caption{Boxplot of evaluation scores in the Only-One (OO) configuration, where only one contextual component is provided at a time. The \textit{current task step} (Model~c) and \textit{hand action} (Model~d) contexts yield the highest median scores (around~4), while \textit{task steps} (Model~b) also supports strong performance. Model~c achieves the slightly higher median, whereas Model~d exhibits greater consistency with fewer extreme outliers, indicating more stable behavior across samples.}
    \label{fig:oo_boxplot}
\end{figure}

In the OO configuration, we provided only one contextual component at a time to evaluate its standalone contribution to model performance. Specifically, Model~a received only the \textit{task duration}, Model~b was given the \textit{task steps}, Model~c had access to the \textit{current task step}, and Model~d utilized only the \textit{hand action data}. This setup isolates the individual effectiveness of each modality, task description, task steps, current task step, and hand actions data, enabling a direct comparison of their relative informativeness.
As illustrated in Figures~\ref{fig:oo_winner} and~\ref{fig:oo_boxplot}, the results show that task description, current task step, and hand action data each provide sufficient context for the model to generate meaningful and relevant responses (higher than $30\%$ performance). 

Notably, task description and current task step yield comparable performance, indicating that both high-level and fine-grained contextual cues independently capture essential task information.
Since current task step and hand action data represent finer-grained, temporally specific context, they lead to stronger performance improvements. Figure~\ref{fig:oo_boxplot} demonstrates that Models~c and~d achieve the highest overall scores, with median ratings around~4. Model~c exhibits slightly higher central tendency but shows more extreme outliers, suggesting occasional failures despite generally strong performance. Model~d, by contrast, displays a slightly wider spread but fewer low-performing cases, indicating greater stability and reliability across samples.
Overall, these findings suggest that while the current task step can produce the highest individual peaks, the hand action context yields the most consistent performance. Therefore, under resource-constrained or unimodal conditions, providing hand action data alone to the LLM assistant can offer the most robust and dependable guidance signal for real-time task understanding.

\section{Discussion}
\label{discussion}
In this study, we introduced a comprehensive framework for context-aware, real-time LLM assistants designed to support task guidance in AR environments. Our findings highlight that fine-grained physical and semantic context—captured through multimodal cues such as hand actions, task steps, and dialogue—substantially enhances the reasoning and response quality of LLM-based assistants. This demonstrates the strong ability of LLMs to act as competent, real-time instructors in AR/VR settings, moving beyond static, pre-scripted guidance toward adaptive, interactive task support. 

The framework’s primary strength lies in its scalability and generality: it can integrate new sensing modalities and prompting strategies without retraining, enabling flexible deployment across various AR/VR systems. However, its current performance still depends on high-quality contextual annotations and structured task representations, which may limit generalization to unstructured environments. Despite these challenges, the study provides clear evidence that contextual grounding allows LLMs to bridge perception and reasoning, offering practical pathways for intelligent, human-centered task assistance.  

\begin{figure*}
  \centering
  \includegraphics[width=\textwidth]{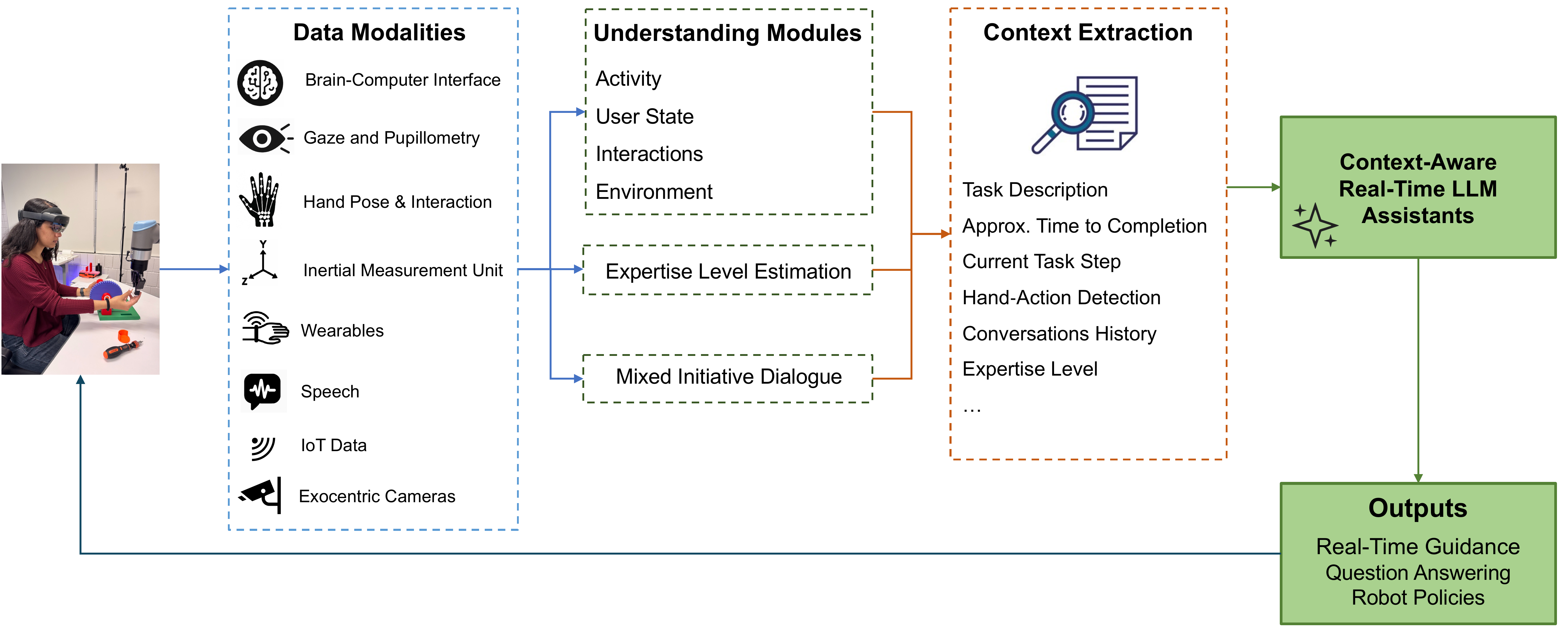}
  \caption{Overview of the proposed context-aware LLM assistant system applied to real-world AR/VR scenarios. The proposed framework in this paper is highlighted in green, illustrating the scope of our current implementation within the broader framework.}
  \label{fig:overview}
\end{figure*}

Yet, several fundamental research directions remain:

\begin{itemize}

    \item \textit{Deployment in real-world AR/VR environments}: Future research should extend this framework to real-world AR and VR task environments. To this end, we plan to collect and annotate a new dataset of human–human collaborative interactions and establish a benchmark for context-aware LLM assistants in human–robot collaboration settings. These will involve more cognitively challenging and physically demanding real-world tasks, such as assembly and maintenance. In collaboration with researchers developing state-of-the-art hand action recognition and step detection models, we will integrate these techniques into live scenarios to extract essential contextual cues required by the assistant in real time (see Figure~\ref{fig:overview}).

    \item \textit{Expansion of modalities and capabilities}: Future studies may also enrich the contextual input space by incorporating additional behavioral signals such as gaze trajectories, ego-/exocentric video streams, and video captions. While the current work focuses primarily on question answering, we aim to expand the LLM assistant’s capabilities toward proactive task guidance—detecting when a user is stuck and offering help without explicit queries—and toward initiating specific robot behaviors in human–robot interaction tasks. Additionally, we plan to extend the framework to support fully immersive VR environments alongside AR.

    \item \textit{Foundation models for behavioral understanding}: A fundamentally challenging and potentially transformative direction is to develop multimodal foundation models (LLM/VLM/VLA) capable of directly interpreting raw behavioral cues (e.g., gaze, gestures, body pose) and physiological signals (e.g., EEG, heart rate variability) to adaptively guide users and proactively intervene during complex task execution. Our current implementation—even when deployed in AR for real-time question answering—relies on intermediate, interpretable inferences for activity understanding (e.g., step recognition, error detection) and user state estimation (e.g., cognitive load, visual attention). Future research should explore end-to-end inference models capable of engaging in mixed-initiative dialogue grounded directly in raw, multimodal data from AR/VR headsets and wearable sensors.

    \item \textit{Privacy, safety, and overreliance on technology}: The proposed framework depends on continuous capture and monitoring of behavioral data, which may compromise user privacy and therefore limit adoption. New approaches are needed to ensure secure, anonymized data collection, processing, and disposal. Moreover, LLMs can make errors—sometimes with serious implications. Thus, it is essential to design safeguards that prevent harmful outcomes and to implement dynamic mechanisms for shared control between human and AI. Additionally, continuous exposure to assistive technologies may lead to overdependence—just like how GPS navigation can limit our ability to navigate independently. To prevent these undesirable consequences, future systems should adaptively scaffold and fade guidance based on user performance, expertise, and interaction history to preserve and strengthen human skills.
\end{itemize}
Overall, this work represents an important step toward building intelligent, responsive, and contextually grounded AR/VR systems enabled by large language models.

\section*{Data Availability}
The data used in this study are publicly available. The multimodal dataset employed in this work is based on the \href{https://holoassist.github.io}{HoloAssist dataset} released by Microsoft Research \cite{Wang2023}. The associated code and sample data for the proposed workflow reproduction are hosted at this GitHub repository: \href{https://github.com/mahyaqorbani/Teaching-LLMs-to-See-and-Guide-Context-Aware-Real-Time-Assistance-in-Augmented-Reality}{https://github.com/mahyaqorbani/Teaching-LLMs-to-See-and-Guide-Context-Aware-Real-Time-Assistance-in-Augmented-Reality}. Researchers may reuse or extend the dataset and code for further applications and benchmarking, provided that they properly cite this work.

\section{Conclusions}
\label{conclusions}
In this work, we presented a novel framework for developing context-aware, real-time LLM assistants capable of providing adaptive task guidance in AR environments. By incrementally integrating diverse contextual inputs, including task description, current task step, hand action data, and dialogue history, our approach systematically examined how multimodal information enhances an LLM’s ability to deliver precise, contextually grounded responses.
Through comprehensive evaluation using lexical comparison, LLM-as-a-judge,  and human evaluation, our results consistently demonstrated that richer contextual grounding leads to significant improvements in accuracy, relevance, completeness, and clarity. Both automated and human evaluations revealed strong alignment, confirming that our proposed LLM-as-a-judge framework offers a reliable, scalable alternative to human evaluation for multimodal task guidance systems. Furthermore, ablation studies underscored the synergistic role of multimodal inputs, with hand action and current task step information emerging as particularly critical for maintaining situational awareness and response consistency.
Overall, these findings highlight the transformative potential of multimodal integration in LLM-driven AR/VR assistance, moving beyond static, pre-scripted guidance toward intelligent, adaptive, and human-like interaction. This study establishes a foundation for future research into proactive, embodied, and contextually aware AI assistants capable of understanding user intent, anticipating needs, and delivering seamless real-time support across industrial, educational, and collaborative domains.
In future work, we plan to extend this framework to live AR/VR deployments, incorporating additional sensory modalities such as gaze, vision, and spatial context, and enabling proactive behavior generation.


\appendices

\section{Models' Prompt Templates}
\label{appendix:prompts}

This appendix presents the full prompt templates used for each model in our incremental prompting framework.  
Each template illustrates the exact structure and contextual inputs supplied to the LLM during inference.  
The placeholders (e.g., \texttt{[task\_type]},\texttt{ [task\_steps]},\texttt{ [task\_duration]}) represent dynamically injected values from the dataset at runtime.  
For clarity, we include the templates for all four models, which differ only in the contextual information they receive.

\begin{tcolorbox}[title=Prompt Template for Model 1 (Approximate Duration of the Task),
colback=gray!5!white, colframe=gray!40!black, fonttitle=\bfseries, breakable]
\small
You are an instructor standing beside a student who is performing the task: \texttt{[task\_type]}.

\medskip
The total duration of the task is approximately \texttt{[task\_duration]} seconds.

\medskip
Here is the conversation history up to \texttt{[current\_time]} seconds:
\begin{quote}
Student: \texttt{[question\_1]} at time \texttt{[t\_1]} seconds \\
Instructor: \texttt{[answer\_1]} \\
Student: \texttt{[question\_2]} at time \texttt{[t\_2]} seconds \\
Instructor: \texttt{[answer\_2]} \\
...
\end{quote}

If there has been no prior question, note that this is the student’s first inquiry.

\medskip
Now, the student asks (at \texttt{[current\_time]} seconds):  
Student: \texttt{[current\_question]}

\medskip
Please respond like a real-time instructor: clear, concise, short, and based only on the context provided.
\end{tcolorbox}

\begin{tcolorbox}[title=Prompt Template for Model 2 (+ Task Steps),
colback=gray!5!white, colframe=gray!40!black, fonttitle=\bfseries, breakable]
\small
You are an instructor standing beside a student who is performing the task: \texttt{[task\_type]}.

\medskip
This task usually follows these common steps (though the exact sequence may vary slightly across different sessions):
\texttt{[task\_steps]}

\medskip
The total duration of the task is approximately \texttt{[task\_duration]} seconds.

\medskip
Here is the conversation history up to \texttt{[current\_time]} seconds:
\begin{quote}
Student: \texttt{[question\_1]} at time \texttt{[t\_1]} seconds \\
Instructor: \texttt{[answer\_1]} \\
Student: \texttt{[question\_2]} at time \texttt{[t\_2]} seconds \\
Instructor: \texttt{[answer\_2]} \\
...
\end{quote}

If there has been no prior question, note that this is the student’s first inquiry.

\medskip
Now, the student asks (at \texttt{[current\_time]} seconds):  
Student: \texttt{[current\_question]}

\medskip
Please respond like a real-time instructor: clear, concise, short, and based only on the context provided.
\end{tcolorbox}

\begin{tcolorbox}[title=Prompt Template for Model 3 (+ Current Task Step),
colback=gray!5!white, colframe=gray!40!black, fonttitle=\bfseries, breakable]
\small
You are an instructor standing beside a student who is performing the task: \texttt{[task\_type]}.

\medskip
This task usually follows these common steps (though the exact sequence may vary slightly across different sessions):
\texttt{[task\_steps]}

\medskip
The total duration of the task is approximately \texttt{[task\_duration]} seconds.

\medskip
Here is the full conversation history up to \texttt{[current\_time]} seconds:
\begin{quote}
Student: \texttt{[question\_1]} at time \texttt{[t\_1]} seconds \\
Instructor: \texttt{[answer\_1]} \\
Student: \texttt{[question\_2]} at time \texttt{[t\_2]} seconds \\
Instructor: \texttt{[answer\_2]} \\
...
\end{quote}

If there has been no prior question, note that this is the student’s first inquiry.

\medskip
The current step of the task is: \texttt{[task\_step]}.

\medskip
Now, the student asks (at \texttt{[current\_time]} seconds):  
Student: \texttt{[current\_question]}

\medskip
Please respond like a real-time instructor: clear, concise, short, and based on the context you have available.
\end{tcolorbox}

\begin{tcolorbox}[title=Prompt Template for Model 4 (+ Hand Action Data),
colback=gray!5!white, colframe=gray!40!black, fonttitle=\bfseries, breakable]
\small
You are an instructor standing beside a student who is performing the task: \texttt{[task\_type]}.

\medskip
This task usually follows these common steps (though the exact sequence may vary slightly across different sessions):
\texttt{[task\_steps]}

\medskip
The total duration of the task is approximately \texttt{[task\_duration]} seconds.

\medskip
Here is the full conversation history up to \texttt{[current\_time]} seconds:
\begin{quote}
Student: \texttt{[question\_1]} at time \texttt{[t\_1]} seconds \\
Instructor: \texttt{[answer\_1]} \\
Student: \texttt{[question\_2]} at time \texttt{[t\_2]} seconds \\
Instructor: \texttt{[answer\_2]} \\
... 
\end{quote}

If there has been no prior question, note that this is the student’s first inquiry.

\medskip
The current step of the task is: \texttt{[task\_step]}.

\medskip
Recent hand actions grouped by task step up to \texttt{[current\_time]} seconds (time of the current question):
(Each action is detected in the form of a verb–noun pair and was extracted using a state-of-the-art hand action recognition model.)
\begin{quote}
Step 1 (\texttt{[step\_name]}): \texttt{[verb\_1 noun\_1]}, \texttt{[verb\_2 noun\_2]}, ... \\
Step 2 (\texttt{[step\_name]}): \texttt{[verb\_3 noun\_3]}, \texttt{[verb\_4 noun\_4]}, \\
...
\end{quote}

\medskip
Now, the student asks (at \texttt{[current\_time]} seconds):  
Student: \texttt{[current\_question]}

\medskip
Please respond like a real-time instructor: clear, concise, short, and based on the context you have. 
Pay close attention to the hand action data, as it offers real-time insights leading up to the moment the question was asked.
\end{tcolorbox}

\section{Prompt for Task Description Extraction}
\label{appendix:task_description_prompt}

Before constructing model inputs, we employed GPT-4 to synthesize unified task descriptions from multiple narrations within the HoloAssist dataset.  
The following prompt was used to aggregate common procedural steps and estimate average task duration across samples.

\tcbset{
  myboxstyle/.style={
    colback=gray!5!white,
    colframe=gray!50!black,
    fonttitle=\bfseries,
    coltitle=white,
    boxrule=0.4pt,
    arc=2mm,
    left=2mm,
    right=2mm,
    top=1mm,
    bottom=1mm,
    breakable
  }
}

\begin{tcolorbox}[title=Prompt Template for Task Description Generation,colback=gray!5!white, colframe=gray!40!black, fonttitle=\bfseries, breakable]
\small
You are provided with several narrations and their corresponding durations (in seconds), each describing how a student performs the task \texttt{[task\_type]}.

\medskip
Your goal is to:

\medskip 
1. Write a clear, concise, and high-level \textbf{task description} summarizing the common steps involved across the narrations.  
   Focus on clarity, avoid repetition, and capture the typical sequence of actions.

\medskip   
2. Calculate and report the \textbf{approximate average duration} of the task, formatted in seconds.

\medskip
\textbf{Output Format:}
\begin{quote}
Task: \texttt{[task\_type]}  
Description:  
[Step-by-step summary of the task]  

Average Duration: [SS] seconds  
\end{quote}

\medskip
Use the following narrations and durations as input:  
\begin{quote}
1.\texttt{[Narration 1]} (\texttt{[Duration 1]}s)  
\\
2.\texttt{[Narration 2]} (\texttt{[Duration 2]}s)  
\\
...  
\end{quote}
\end{tcolorbox}

\section{LLM-as-a-judge Prompt Template}
\label{appendix:judge_prompt}

The following template was used to evaluate model responses using an LLM-as-a-judge.  
The model was given all contextual elements available up to the time of each student question and instructed to rate the responses on multiple dimensions.

\newpage
\begin{tcolorbox}[title={LLM-as-a-Judge Prompt Template},,
colback=gray!5!white, colframe=gray!40!black, fonttitle=\bfseries, breakable]
\small
You are an expert judge evaluating and comparing four AI-generated responses to a student's question during a technical task.

\medskip
---\\
Task Type: \texttt{[task\_type]}

\medskip
Total Task Duration: \texttt{[task\_duration]} seconds  

\medskip
Task Steps (may vary by session): \texttt{[task\_steps]}

\medskip
Time of Student Question: \texttt{[current\_time]} seconds  

\medskip
Conversation History:  
\texttt{[previous student\_instructor exchanges, if any]}

\medskip
Current detected task step: \texttt{[task\_step]}

\medskip
Hand actions observed earlier in the task (start → last question):  
\texttt{[earlier\_hand\_actions\_data]}

\medskip
Most recent hand actions (last question → now \texttt{[current\_time]}s):  
\texttt{[recent\_hand\_actions\_data]}

---\\

Current Student Question (asked at \texttt{[current\_time]} seconds):  
\texttt{[current\_question]}

\medskip
Reference instructor reply (may be imperfect; do NOT require a match):  
\texttt{[instructor\_answer]}

AI Responses:  \\
- Model 1: \texttt{[response\_1]}\\
- Model 2: \texttt{[response\_2]}\\
- Model 3: \texttt{[response\_3]}\\
- Model 4: \texttt{[response\_4]}\\

---\\

\# Evaluation Instructions\\

Judge the responses using ONLY the shared context above.  
The order of candidates may be randomized; ignore position and verbosity.

\medskip
Score each model on:
\begin{itemize}
\item Correctness: 0–5 (5 = fully accurate; 3 = minor issues; 1 = incorrect; 0 = wrong)
\item Completeness: 0–5 (5 = fully answers; 3 = partial; 1 = minimal; 0 = none)
\item Contextual Relevance: 0–5 (5 = fully grounded in the provided context/hand actions; 3 = weakly grounded; 1 = irrelevant; 0 = off-topic)
\item Clarity: 0–5 (5 = very clear; 3 = somewhat clear; 1 = vague; 0 = incomprehensible)
\item Final Score: 0–5 (holistic judgment; NOT an average)
\end{itemize}

\medskip
Return ONLY the following format for each model:

\medskip
Model: Model N\\
Correctness: X/5\\
Completeness: X/5\\
Contextual Relevance: X/5\\
Clarity: X/5\\
Final Score: X/5

\medskip
Do not include per-model explanations.  
At the end, add a short comparison summary (1–3 sentences) identifying which model(s) performed best and why.

\medskip
Comparison Summary: ...

\end{tcolorbox}

\appendices

\end{document}